\documentclass[aps,pre,graphicx,reprint,amssymb,showpacs,superscriptaddress,floatfix,nobalancelastpage]{revtex4-1}

\usepackage[T1]{fontenc}
\usepackage{amsmath,amsfonts}
\usepackage{amssymb,textcomp}
\usepackage[xcdraw]{xcolor}
\usepackage{tikz}
\usepackage{bm}
\usepackage[bookmarks=false]{hyperref}
\usepackage{graphicx, wrapfig}
\usepackage[caption=false]{subfig}
\usepackage{natbib}
\usepackage{comment}

\newcommand\beq{\begin{equation}}      % begin equation numbered
\newcommand\beqnn{\begin{eqnarray*}}   % begin equation no number
\newcommand\beqa{\begin{eqnarray}}     % begin equation array
\newcommand\beqann{\begin{eqnarray*}}  % begin equation array no number

\newcommand\eeq{\end{equation}}        % end equation numbered
\newcommand\eeqnn{\end{eqnarray*}}     % end equation no number
\newcommand\eeqa{\end{eqnarray}}       % end equation array numbered
\newcommand\eeqann{\end{eqnarray*}}    % end equation array no number

\newcommand{\ave}[1]{\langle #1 \rangle}
\def\nl {\nonumber \\}

\newcommand\bi{\begin{itemize}}
\newcommand\ei{\end{itemize}}

\def\nl {\nonumber \\}

\newcommand{\eref}[1]{(\ref{#1})}
\newcommand{\sref}[1]{Section~\ref{#1}}
\newcommand{\fref}[1]{Fig.~\ref{#1}}

\newcommand{\al}[1]{\begin{align} #1 \end{align}}
\def\l0{L}
\def\txz{T_{xz}^\phi}
\def\czzp{C(z,z',\xi)}

\def\tt{T_{tot}}
\def\ha{\text{H(A) }}

\def\fp{f_\phi}
\def\fv{\bm f_{\bm v}}
\def\czzpl{C^\Lambda (z,z',\xi)}
\def\czzplr{\tilde C^{\tilde \Lambda} (\tilde z,\tilde z',\tilde \xi)}

\def\czzplxiinfr{\tilde C^{\tilde \Lambda} ( \tilde z,\tilde z',\tilde \xi=\infty)}
\def\D{\mathcal D}

\def\cm{ }%\def\cm{\checkmark}

\def\etaeff{\eta_{\textrm{eff}}}

\begin{document}

\preprint{[Draft V1]}

\title{
Viscosity of a sheared correlated (near-critical) model fluid in confinement}

\author{Christian M. Rohwer}
\email[]{crohwer@is.mpg.de}

\affiliation{Max Planck Institute for Intelligent Systems, Heisenbergstr.~3, 70569 Stuttgart, Germany}
\affiliation{4th Institute for Theoretical Physics, Universit\"at Stuttgart, Pfaffenwaldring 57, 70569 Stuttgart, Germany}

\author{Andrea Gambassi}
\affiliation{SISSA --- International School for Advanced Studies and INFN, via Bonomea 265, I-34136 Trieste, Italy}

\author{Matthias Kr\"uger}
\affiliation{Max Planck Institute for Intelligent Systems, Heisenbergstr.~3, 70569 Stuttgart, Germany}
\affiliation{4th Institute for Theoretical Physics, Universit\"at Stuttgart, Pfaffenwaldring 57, 70569 Stuttgart, Germany}

\date{\today}

\begin{abstract}
Second-order phase transitions are characterized by a divergence of the spatial correlation length of the order parameter fluctuations. For confined systems, this is known to lead to remarkable equilibrium physical phenomena, including finite-size effects and critical Casimir forces. We explore here some non-equilibrium aspects of these effects in the stationary state resulting from the action of external forces: by analyzing a model of a correlated fluid under shear, spatially confined by two parallel plates, we study the resulting viscosity within the setting of (Gaussian) Landau-Ginzburg theory. Specifically, we introduce a model in which the hydrodynamic velocity field (obeying the Stokes equation) is coupled to an order parameter with dissipative dynamics. The well-known Green-Kubo relation for bulk systems is generalized for confined systems. This is shown to result in a non-local Stokes equation for the fluid flow, due to the correlated fluctuations. The resulting effective shear viscosity shows universal as well as non-universal contributions, which we study in detail. In particular, the deviation from the bulk behavior is universal, depending on the ratio of the correlation length and the film thickness $L$. In addition, at the critical point the viscosity is proportional to $\ell/L$, where $\ell$ is a dynamic length scale. These findings are expected to be experimentally observable, especially for systems where the bulk viscosity is affected by critical fluctuations.

\end{abstract}

\pacs{74.40.Gh, 47.27.N-, 64.60.De}

\maketitle 

\section{Introduction}

\label{sec:intro}

Correlations in thermodynamic systems can arise for a variety of reasons. Examples include systems with intrinsic length or time scales (e.g., colloidal dispersions, polymers or biological fluids) or those {in which} 
large-scale fluctuations emerge due to {the occurrence of} second order phase transitions {of various nature}. 
Spatial confinement of these systems can lead to {novel} 
physical {phenomena affecting the behavior of soft matter}, 
ranging from entropic or depletion forces in polymer systems to
thermodynamic (or critical) Casimir forces mediated by near-critical fluids. 
These forces are {the} classical analogue %s 
of the quantum Casimir forces which occur when the fluctuations of the quantized electromagnetic field are spatially confined \cite{bordag,krechbook,dantchevbook,gambassi2008-rev}. 

Properties of confined systems have been studied extensively, primarily \textit{at thermodynamic equilibrium}. In this setting, the possible emergence of collective behaviours, with the associated universality and scaling laws, allows the introduction of simplified minimal models which form the basis of theoretical descriptions applicable to a large class of physical systems, belonging to the same so-called universality class.
{Examples of these studies include finite-size scaling, wetting phenomena \cite{Nightingale1985prl}, effective forces in critical films with various homogeneous \cite{krechdietrich1991prl}, rough \cite{kardarliPRL1991} or chemically patterned 
surfaces \cite{nellen2009,soyka2008,gambassi2011critical,trondle2011}, as well as many-body effects \cite{tuna2016}.}

{These effective forces have recently been the subject of very detailed experimental investigations, both in the case of quantum \cite{bordag} and critical \cite{hertlein2008,gambassi2008-rev} confined fluctuations. Such fluctuation-induced interactions facilitate the experimental manipulation of colloidal aggregations through correlated solvents, as they provide an exquisite experimental control over the range and magnitude of colloidal interactions through temperature 
changes \cite{bechingerarcher,schall,gambassicomment2010prl,faber2013,shelke2013,veen2012}, which may find practical applications in soft matter and beyond  \cite{marino2016}.}
Correlated systems also exhibit non-trivial \textit{non-equilibrium properties}. {In the bulk,} 
various transport coefficients {of fluid media,} such as the 
viscosity, are sensitive to {the occurrence} of phase transitions \cite{onukibook}.
{Phase separation, in turn, can be affected by the presence of strong shear \cite{corberi1998,corberi1999}.}
Due to the subtle interplay between fluctuations at different time- and length scales, spatially confined systems are 
expected to display a wealth of additional phenomena out of equilibrium {in addition to interesting dynamical properties at equilibrium}. Some of them have been investigated for film geometries, where time-dependent linear response and correlation functions 
{were studied} for a fluctuating medium with purely dissipative dynamics (the so-called Model A \cite{hohenberg}) in Refs.~\cite{gambassi2006,gambassi2008EPJB}, {while the} %The 
behavior of critical Casimir forces away from equilibrium was investigated in Refs.~\cite{gambassi2008EPJB,deangopinathan2010PRE,hanke,gambassi2013prl}; {drag forces, instead, were studied for inclusions moving within a medium \cite{demery2010,gambassi2013prl}}. Non-equilibrium fluctuations arising from conservation laws have also been demonstrated to lead to Casimir forces far from criticality \cite{kirkpatricksengers2013,aminovkardarkafri2015,rohwer2017transient}.  
In this context the role of the stress tensor has been discussed in detail  \cite{deangopinathan2009JStatMech,deangopinathan2010PRE,bitbolfournier2011forces}. 
State-of-the-art experimental methods now allow extremely fine and accurate measurements of the dynamic properties of fluid systems not only in the bulk (e.g., the viscosity of helium near the critical point \cite{agosta}), but also in confinement, e.g., the frictional parameters for confined complex fluids \cite{israelashvili1996} and the dynamic shear viscosity of colloidal suspensions with varying shear \cite{cheng2011,brader2010}.

Theoretically, the behavior of sheared fluids in confinement away from phase transitions (i.e., in the absence of long-range correlations within the fluid solvent) has been studied with various approaches \cite{bocquetbarrat,harrowell,aerov2014,aerov2015}. However, to our knowledge, the combined effects of shearing and correlated fluctuations on the physical properties of a confined systems have not {yet} been investigated, in spite of the fact that they are within the reach of current experimental techniques.   

In this work we study the {viscosity of a confined, correlated fluctuating medium}, such as a fluid, via linear response theory. In order to highlight the relevant effects of correlated fluctuations, we focus on a simple dynamical model,
which is a modified version of the so-called dynamical Model H \cite{hohenberg} for binary liquid mixtures. We analyze the effective viscosity $\eta_{\rm eff}$ of the fluid as a function of the bulk correlation length $\xi$ of the fluctuations of its order parameter and as a function of the separation $L$ of the parallel surfaces which confine the medium within a film. Specifically, we shall assume that the dynamics of the order parameter is dissipative, coupled to a hydrodynamic flow field obeying Stokes equation. 
While the resulting stress and viscosity from this model
are cutoff-dependent, i.e., they depend on microscopic parameters of the fluid, their dependence on the correlation length $\xi$ is to some extent universal. 
{The viscosity $\eta_{\rm eff}$ turns out to depend on the ratio between two length scales, i.e., the one determined by the dynamical parameters of the model and the film thickness $L$}.
Further scaling appears through a dimensionless function of $L/\xi$.

The rest of the presentation is organized as follows: in \sref{syssect} we introduce the {fluid} system under investigation. In \sref{coarsegrainsect} we set out the dynamical model we focus on, which accounts for the coupling to hydrodynamics as in Model H. We show that the model satisfies the so-called potential conditions, which ensure that, in the absence of shear, the system relaxes to the correct equilibrium distribution. We then discuss the use of the stress tensor in non-equilibrium conditions.
In \sref{linrespsect}, we derive Green-Kubo relations for the system in confinement, which allow us to define the viscosity of the {confined} fluid. Then we show that a non-local viscosity naturally arises as a consequence of the correlated fluctuations.  ~\sref{resultssect} presents explicit expressions of the viscosity for a Gaussian Hamiltonian, with the main prediction given in Eq.~\eqref{etaeffminbulk}.
We close with a discussion of our results and an outlook in \sref{discussionsect}.

\section{The system}
\label{syssect}

\begin{figure}[t]
 \begin{center}
 \includegraphics[width=0.45\textwidth]{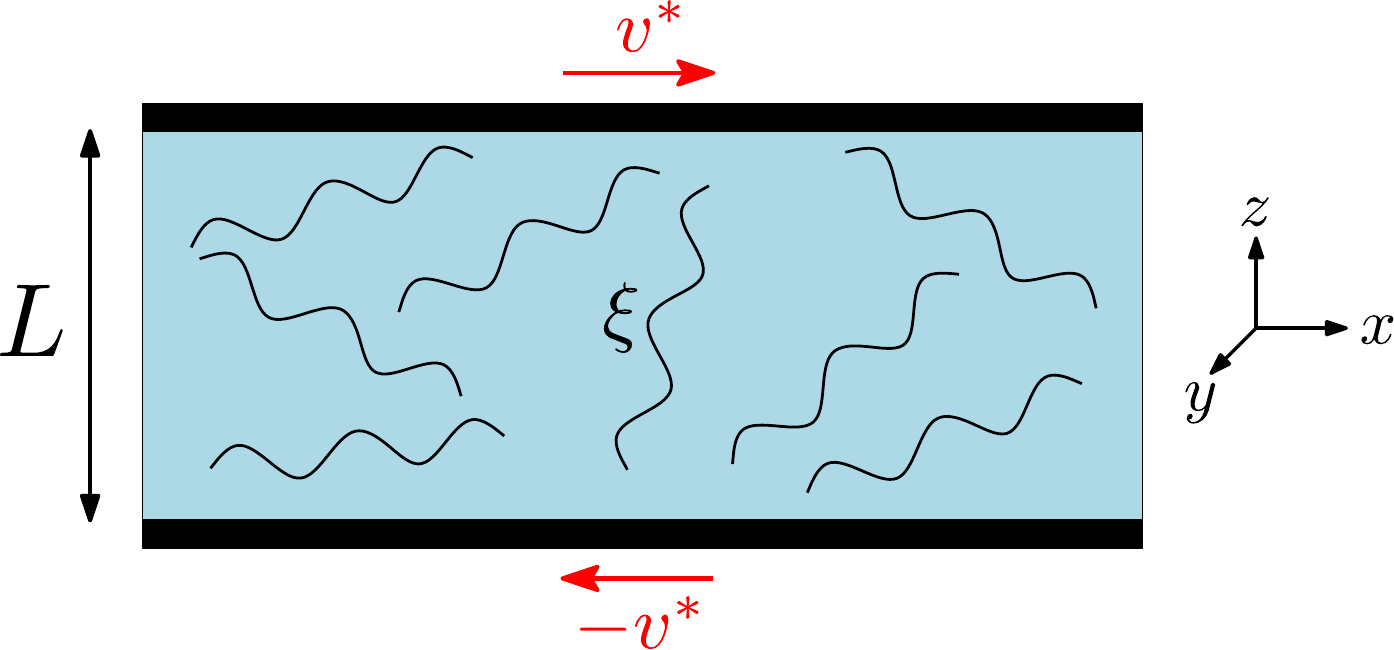}
 \end{center}
 \caption{Sketch of a %complex 
 fluid {characterized by fluctuations (represented by wiggly lines)} with bulk correlation length $\xi$, confined within a film of thickness $L$ {(and large transverse area $A$)} %and 
subject to the shear induced by the velocity $v^*$ imposed to the boundaries, as indicated by the red arrows. The force necessary to keep these surfaces in steady motion defines a distance-dependent effective viscosity, which we investigate here.}
 \label{fig:confinedshear}
\end{figure}

We consider here a soft system, such as a fluid, with a correlation length $\xi$ characterising the spatial correlation of fluctuations  \cite{onukibook,kardarbook}. The system is spatially  confined along  the $z$-direction by two planar plates positioned at $z= \pm L/2$, so that the medium occupies the domain $z \in  \D = [-L/2,L/2]$, being infinitely extended {along} the $x$-$y$ plane, as shown in \fref{fig:confinedshear}. 
We assume that any microscopic or molecular length scale $a$ of the fluid is small compared to both $\xi$ and $L$. These conditions are met, for example, in fluids close to second-order phase transitions. 
In order to investigate the effect of shear, the confining plates are set into relative motion along the $x$ direction with velocities $\pm v^*$, resulting in a velocity difference {between the upper and lower plate} of $2|v^*|$.
%
%
%%%%%%%%%%%%%%%%%%%%%%%%%%%
%%%%%%%%%%%%%%%%%%%%%%%%%%%
%
%
Here {we assume} $v^*$  
{to be} time-independent, and the system 
to {have reached} 
a steady state. 
The bare shear rate is denoted by $s_0\equiv 2|v^*|/L$. 
It is also assumed that $v^*$ is sufficiently small so that non-linear effects in {the fluid response} 
to shear are negligible. In other words, we focus on the linear response regime, in order to keep the discussion as simple as possible.
Within this setting, the main quantity of interest is the force (per area $A$) {which has to act on the plates in order to keep them in steady} 
motion at the given velocities $\pm v^*$, and the associated effective viscosity 
\begin{align}
\eta_{\rm eff}=\frac{|\Delta F_x| }{A s_0},
\label{eta}
\end{align}
{where $\Delta F_x $ is the net force difference between the two plates.} $\Delta F_x $ may also be seen as the force necessary to move the upper plate with velocity $2 v^*$, if the lower is at rest.
In case the correlation length is small compared to $L$, $\eta_{\rm eff}$ is expected to approach the bulk viscosity of the fluid. For larger $\xi$, it is natural to expect that $\eta_{\rm eff}$ will eventually develop a dependence on $L$. In order to study the dependence of $\eta_{\rm eff}$ on $L$ and $\xi$, the first step is to determine the (coarse-grained) velocity profile $v_x(z)$ for a given $\xi$, which will be shown to be the solution of a {\it non-local} Stokes equation in Sec.~\ref{stokessect}. From it, $\eta_{\rm eff}$ can finally be determined.

\section{A coarse-grained model for correlated fluids}
\label{coarsegrainsect}

This section outlines the {model and discusses its various ingredients.} 
First we discuss the coarse-grained description {of the fluid which accounts for} 
the relevant fluctuation phenomena in terms of an order parameter, {introducing a suitable (simplified) dynamics and commenting on the role of the} dynamical conservation laws. We then address how 
{this} dynamics may be coupled to a hydrodynamic velocity field, {which is an essential ingredient of fluid systems and is necessary in order to describe shear forces acting on the plates in Fig.~\ref{fig:confinedshear} as detailed in Sec.~\ref{fsect}.}

\subsection{Models for statics and dynamics}
\label{modABsect}

In order to explore how correlated fluctuations influence the viscosity under confinement, we consider a minimal model in which the relevant physical properties of the 
system can be effectively described in terms of an order parameter $\phi(\bm r,t)$ and of the corresponding Landau-Ginzburg effective free energy $\mathcal H[\phi]$. The latter controls the equilibrium distribution $\mathcal P_e[\phi] \propto \exp(-\mathcal H[\phi]/(k_BT))$ of the fluctuations of $\phi$ at temperature $T$, where $k_B$ is Boltzmann's constant.
{The effective free energy} $\mathcal H[\phi]$ arises from coarse-graining to length scales which are large compared to the microscopic scale $a$, which yields a continuum description in terms of the field  $\phi(\bm r,t)$. The form of $\mathcal H[\phi]$ is obtained from an expansion in powers of the order parameter and its derivatives, ordered according to their relevance (in the sense of renormalization-group theory, see, e.g., Ref.~ \cite{kardarbook}),
which {describes} 
the relevant {physical properties of the system} 
if the spatial correlation length $\xi$ is macroscopically large,
\al{
\mathcal H [\phi]= \int d\bm r \; H[\phi],
\label{H1}
}
with
\al{
H[\phi] = \frac 1 2  (\nabla \phi)^2  + \frac  r 2 \phi^2(\bm r)+ \frac{u}{4!} \phi^4(\bm r) +\ldots \;\cm,
\label{H1b}
}
where $u>0$. The parameter $r$ 
controls the correlation length $\xi$ which, in the absence of confinement, formally diverges as $r$ approaches a critical value $r_c$. 
This kind of effective Hamiltonian $\mathcal H [\phi]$ captures universal phenomena associated with second-order phase transitions in a vast class of physical systems, see, e.g., Refs.~ \cite{hohenberg,onukibook,kardarbook}.
In these cases, one assumes a simple temperature dependence of $r$ so that $r$ approaches $r_c$ when $T$ approaches the critical temperature $T_c$ of the phase transition. 
An example is the demixing transition of binary mixtures of fluids, in which the order parameter field $\phi$ is proportional to the deviation of the local concentration of one of the two components from the bulk concentration \cite{onukibook}.

The quantitative description of critical phenomena for $T\to T_c$ requires considering the effects of non-Gaussian fluctuations of $\phi$, %i.e., 
i.e., $u\neq 0$ in Eq.~\eqref{H1b}. 
However, for our purposes, interesting qualitative effects emerge already in the {simpler} case of Gaussian fluctuations with $u=0$, which we focus on below. 
Correspondingly, $r_c=0$ and the (bulk) correlation length $\xi$ is related to $r \ge 0$ in Eq.~\eqref{H1b} as $\xi = r^{-1/2}$.
Boundaries like the two plates in Fig.~\ref{fig:confinedshear} can either be described by introducing surface terms in the Hamiltonian $\mathcal H [\phi]$ or by imposing boundary conditions on $\phi$ \cite{diehl-book}. 
As we focus below on the case $u=0$, the latter approach is more convenient, and therefore we assume that these boundary conditions are of the Dirichlet type, i.e., $\phi=0$ at the surfaces.  %boundary. 
For a binary liquid mixture, for instance, these can be effectively realized by considering chemically patterned substrates \cite{gambassi2011critical}. 

Since the effective free energy in Eq.~\eqref{H1} results from integrating out microscopic degrees of freedom corresponding to a certain mesoscopic configuration of  $\phi$, there is no direct recipe (such as Hamilton's equations of motion of classical mechanics) to obtain the system's dynamics. Accordingly, {an effective} 
dynamical description is typically formulated in terms of Langevin equations for 
$\phi$  \cite{hohenberg,onukibook,kardarbook}, 
\al{
\partial_t \phi(\bm r,t) &=  -\hat \mu \frac{\delta \mathcal H}{\delta \phi} + \theta(\bm r,t), \cm   \label{modAB}
}
where $\hat \mu$ is a mobility (operator), which relates the field force $\delta \mathcal H/\delta \phi$ to the time derivative of $\phi$. The stochastic force $\theta(\bm r,t)$  is chosen such that Eq.~\eqref{modAB} fulfils the fluctuation-dissipation theorem, and the equilibrium fluctuations resulting from  %according to 
Eq.~\eqref{modAB} are distributed according to  $\mathcal P_e[\phi]$, as discussed above. 
In the simplest case, $\theta(\bm r,t)$ is expected to have short-ranged correlations in both time and space, i.e., 
\al{
\ave{\theta(\bm r,t)\theta(\bm r',t')} &= 2 \,\hat \mu\, k_B T \,\delta^{(3)}(\bm r - \bm r') \delta(t-t').
\label{modAB2}
}
{Depending on the conservation laws 
for the field $\phi$ in Eq.~\eqref{modAB}, one distinguishes two important classes of dynamical models: (a) in the absence of a local conservation of $\phi$, i.e., with purely dissipative dynamics (Model A in the notion of Ref.~ \cite{hohenberg}), $\hat \mu = \mu = \mbox{const}$; 
(b) if, instead, $\phi$ satisfies a local continuity equation $\partial_t \phi(\bm r,t) +\bm \nabla \cdot \bm J_\phi(\bm r,t) = 0$ with a suitable stochastic current $\bm J_\phi(\bm r,t)$, then $\phi$ is conserved (Model B in the notion of Ref.~ \cite{hohenberg}) and $\hat \mu =-\mu\nabla^2 \cm$.}
These two models have been studied extensively for bulk systems in the literature, and, to a lesser extent, in the presence of surfaces. In particular, Model B dynamics in a semi-infinite geometry is discussed in
Refs.~ \cite{dietrichdiehl1983,diehljanssen1992,diehl1994,diehlwichmann1995}. Investigation of  dynamical properties near criticality (e.g., of the relaxation {towards} equilibrium) within a film geometry appears to have been limited to dissipative dynamics (see, e.g., Refs.~ \cite{gambassi2006,gambassi2008EPJB,deangopinathan2009JStatMech,deangopinathan2010PRE}), 
due to the additional complexity of boundary conditions arising from the higher derivatives in Model B.
The dynamics specified by Eqs.~\eqref{modAB} and \eqref{modAB2} does not account for the presence of velocity fields, i.e., hydrodynamics, which is an essential ingredient 
when describing shear. We will introduce it in the next subsection.

\subsection{Model H: Coupling to hydrodynamics}
\label{sec:modH}

A minimal model {which captures} %capturing 
the coupling of the order parameter $\phi(\bm r, t)$ to hydrodynamics is the so-called Model H \cite{hohenberg,onukibook,kardarbook}. In particular, $\phi$ is coupled with the velocity field $\bm v (\bm r,t)$ {of the fluid medium}, such that Eq.~\eqref{modAB} is {supplemented by} 
a typical advection term,
\al{
\partial_t \phi(\bm r,t) &= - \bm \nabla \cdot (\phi \bm v) -\hat \mu \frac{\delta \mathcal H}{\delta \phi} + \theta(\bm r,t) . \cm \label{langevinphi} 
}
{In turn, the dynamics of the velocity field $\bm v (\bm r,t)$ follows the 
Stokes equation of a Newtonian, incompressible fluid with hydrodynamic viscosity $\eta_0$, which is affected by the presence of $\phi$,}
\al{
\rho \,\partial_t \bm v(\bm r,t) &= - \,\left(\phi \bm \nabla \frac{\delta}{\delta \phi} \mathcal H\right)_\perp +\eta_0 \,\nabla ^2 \bm v + \bm \zeta _\perp(\bm r,t).\cm \label{langevinv}
}
In Eqs.~\eqref{langevinphi} and \eqref{langevinv}, $\hat \mu$ is the mobility coefficient from Sec.~\ref{modABsect}
%, 
%
%
and $\rho$ is the mass density of the fluid. As the fluid is assumed to be incompressible, $\bm \nabla \cdot \bm v(\bm r,t) = 0$, $\rho$ is spatially homogeneous.
The subscript $\perp$ in Eq.~\eref{langevinv} indicates a projection onto the transverse component of the wave vector in 
Fourier space  \cite{onukibook}, which renders all the terms in Eq.~\eref{langevinv} divergence-free. In the geometrical setup of shear, as considered further below, and within the approximations we shall introduce in Sec.~\ref{linrespsect}, this projection will be automatically implemented.

As before, the stochastic forces in Eqs.~\eqref{langevinphi} and \eqref{langevinv} are chosen to satisfy the fluctuation-dissipation theorem, i.e., additionally to Eq.~\eqref{modAB2},
\al{
\ave{\zeta_i(\bm r,t)\zeta_j(\bm r',t')} &= -2\,\eta_0\,  k_B T \,\delta_{ij}\nl & \qquad\times \nabla ^2\delta^{(3)}(\bm r - \bm r') \delta(t-t')\cm. \label{noisev} 
}
{This choice ensures} 
that the probability distribution of the variables $\bm v$ and $\phi$ relaxes to the Boltzmann equilibrium distribution $\mathcal P_e [\phi,\bm v]$, where the Hamiltonian in Eq.~\eqref{H1} picks up an additional kinetic term from $\bm v$, i.e.,
\al{
\mathcal P_e [\phi,\bm v] \propto \exp\left\{-\beta\left( \mathcal H[\phi] + \frac 1 2 \int d\bm r \rho \bm v^2\right)\right\}.\cm
\label{Peq}
}
{This non-trivial fact follows from the so-called potential 
conditions \cite{onukibook} which fix the form of Eqs.~\eqref{langevinphi} and \eqref{langevinv}.} 
{Indeed, upon adding to  Eq.~\eqref{modAB} the advection term $-\bm \nabla \cdot (\phi \bm v)$ as in Eq.~\eref{langevinphi}, the first term on the r.h.s.~of Eq.~\eqref{langevinv} has to be added in order to fulfil these potential conditions; Additional details are provided in Appendix \ref{potcondsect}.}

Equations~\eqref{langevinphi} and  \eqref{langevinv} thus describe an order parameter $\phi$ with either conserved ($\hat \mu =-\mu\nabla^2 \cm$) or non-conserved ($\hat \mu = \mu = \mbox{const}$) dynamics which is additionally coupled to a hydrodynamic velocity field $\bm v$ and which relaxes towards the equilibrium state in Eq.~\eqref{Peq}. It is important to note that the potential conditions hold in {both} these cases --- see Appendix \ref{potcondsect}.
The former provides, for instance, a suitable description of a binary liquid mixture, since both the order parameter and the fluid momentum are conserved, and is termed Model H in the notion of 
Ref.~ \cite{hohenberg}. 
{The latter, instead, describes a field with purely dissipative dynamics such as colloidal 
particles carrying Ising-spins (see, e.g., Refs.~  \cite{maretkeim2005colloidal,maretkeim2009experimental,maretkeim2009ultrafast}), because the corresponding ``magnetization'' is not necessarily conserved. Equation \eqref{langevinv} then decribes advection by a flow, mimicking the motion of the solvent. A timely example may be given by ferrofluids, i.e., suspensions of magnetized colloids. Here, the nature of possible phase transitions is still debated  \cite{zhang1994global,mamiya2000phase,hansen2002critical,soto2010magnetoviscosity}, and driving or shearing might add additional insights. In what follows we refer to the latter model as Model H(A). 

Despite the fact that the applicability of this model (as compared to Model H) for describing actual physical systems may be more limited, Model H(A) is amenable to simpler analytic calculations in confined systems, as it does not introduce the additional complications at boundaries (which make Model B dynamics a particularly challenging problem). This is the reason why we focus below on Model H(A). }

\subsection{Dimensional analysis of Model H(A)}

\label{dimensionalsect}

Having in mind phenomena appearing on large length scales, it is instructive to investigate the behaviour of the terms in Eqs.~\eref{langevinphi} and \eref{langevinv} upon coarse-graining space and time.  We therefore consider a rescaling transformation \cite{kardarbook}
\al{
\bm r \to b \bm r, \quad t \to b^{z} t, \quad \phi \to b^\chi \phi,\quad \bm v \to b^\psi \bm v.\label{eq:rs}
}
In $d$ spatial dimensions, standard dimensional analysis (see, e.g., Ref.~\cite{kardarbook}) yields the exponents $z=2$, $\chi = 1-d/2$ and $\psi = -d/2$. It follows that the coupling terms $\bm \nabla \cdot (\phi \bm v)$ and $\left(\phi \bm \nabla \delta \mathcal H/\delta \phi \right)_\perp$ in Eqs.~\eref{langevinphi} and \eref{langevinv}, respectively, scale as $\propto b^y$, with the exponent $y = 1-d/2$. Under such coarse-graining transformations, the coupling terms therefore become increasingly irrelevant for $d>2$. However, as the effect investigated below depends on the presence of such couplings, they cannot be set to zero from the outset.

In the considered setup with external shear-driving, we can estimate the importance of the mean value $\propto s_0 z$ of the velocity, see Fig.~\ref{fig:confinedshear} and the corresponding text. Assuming the shear strain not to be affected by the rescaling, we find
\al{
s_0 z \to b^{1-z} s_0z = b^{-1} s_0 z,
}
because $z\to b z$ and $s_0$ is the time derivative of strain. Accordingly, under rescaling,  $s_0 z$ will be more relevant than the fluctuations of $\bm v$ (recall $\psi=-3/2$ in Eq.~\eqref{eq:rs}) in $d=3$, and remains relevant in Eq.~\eref{langevinphi}. This insight will be used   below in Sec.~\ref{linrespsect} in order to simplify our analysis.

\subsection{Stress tensor and forces}
\label{fsect}

The effective Hamiltonian in Eq.~\eqref{H1} is designed to yield the (free) energy of a certain configuration of the mesoscopic field $\phi$ {\it in equilibrium}. 
Away from equilibrium, instead, additional assumptions have to be made, e.g., the {one} %assumption 
that the microscopic degrees of freedom remain (instantaneously) equilibrated and that they exert the same force as they would do in equilibrium (e.g., on boundaries or inclusions). In equilibrium, forces such as  pressure or fluctuation-induced 
forces acting on 
{immersed} objects, can equivalently be 
{calculated} from the free energy  
or via the stress tensor. 
However, out of equilibrium \cite{gambassi2008EPJB} 
additional conceptual issues arise regarding use of the stress tensor \cite{deangopinathan2010PRE}.

The effective viscosity in the setup described in Fig.~\ref{fig:confinedshear} is given in terms of $\Delta F_x$. As the surfaces in Fig.~\ref{fig:confinedshear} are smooth so that the boundary conditions {are} translationally invariant, the Hamiltonian in Eq.~\eqref{H1} is invariant under translations of either plate along $x$.  Accordingly,
the derivative with respect to a (marked) {position $x_p$ along $x$ of one of the two surfaces vanishes identically at all times, i.e., $\partial \mathcal H [\phi]/\partial x_p=0$ and 
the order parameter field $\phi$, together with its Hamiltonian \eqref{H1}, 
can not exert any force along $x$ on the plates. }

Shear forces are 
{therefore} solely transported through the coupled fluid in Model H {and H(A)}. 
The term $\phi \bm \nabla \delta\mathcal H/\delta \phi$ in Eq.~\eqref{langevinv} 
{(due to the potential conditions, see Appendix \ref{potcondsect})}
can be identified with the stress tensor $T^\phi$ corresponding to the field $\phi$, as set out in Ref.  \cite{onukibook}, i.e.,
\begin{align}
-\phi \bm \nabla \frac{\delta\mathcal H}{\delta \phi}  = -\bm \nabla \cdot T^\phi.
\label{fvequalsTphi}
\end{align}
At equilibrium, $T^\phi_{ij} = \delta_{ij} H - \partial_i ( \partial H/\partial_j \phi(\bm r))$ with $i,j\in \{x,y,z\}$. Out of equilibrium, the stress tensor may pick up additional contributions 
(see, e.g., Ref.~  \cite{deangopinathan2010PRE}); however this has no bearing on the off-diagonal components $\txz$ which we require here in order to study the effective viscosity \cite{onukibook}. Therefore the possible issues mentioned at the beginning of the section are inconsequential. 
We may thus reformulate Eq.~\eqref{langevinv} as
\al{
\rho \,\partial_t \bm v(\bm r,t) &= - \,\big(\bm \nabla \cdot T^\phi)_\perp +\eta_0 \,\nabla ^2 \bm v + \bm \zeta _\perp(\bm r,t).\cm \label{langevinvst}
}
The physical interpretation of this equation is now apparent, as the evolution of the velocity field is due to both hydrodynamic viscous (stress-) forces $\sim\nabla ^2 \bm v$ and (stress-) forces arising from the order parameter field $\phi$ via $T^\phi$.
The influence of the moving plates will therefore be transported by the velocity field $\bm v$, which we assume to obey stick boundary conditions at the walls, {which thereby play} %thereby playing 
the role of an anchor for the field $\phi$ with respect to the moving plates. In a steady state, Eq.~\eqref{langevinvst} represents a force balance between viscous forces from the fluid and stresses from the order parameter field. The force acting on the moving walls can thus be determined unambiguously. 

Next, in Section~\ref{linrespsect}, we provide the general solution of Eqs.~\eqref{langevinphi} and \eqref{langevinv} for small shear velocities within linear response theory.

\section{Linear response relations for shear in confined systems}
\label{linrespsect}

Within the film represented in Fig.~\ref{fig:confinedshear}, the stationary mean velocity field $\left\langle{\bm v}(\bm r)\right\rangle_s$  
is expected to depend only on $z$, to point in $\hat x$ direction, and to vanish for $z=0$, due to the symmetries of the problem,  \begin{align}\label{vxdef}\left\langle{\bm v}(\bm r)\right\rangle_s= \hat{x}\left\langle{v}(z)\right\rangle_s,\end{align} where $\langle \ldots \rangle_s$ indicates the average calculated in the steady state.  
We % have 
also introduce the shear rate,
\al{
s(z) = d v(z)/d z,
\label{eq:def-s}
}
which, in contrast to bulk shear, may depend on $z$ in this inhomogeneous system. 
The boundary conditions for %the field 
$v$ can be expressed as
\al{
v(z=\pm L/2)= \int_0^{\pm L/2} dz'\;s(z') = \pm v^*, 
\label{vboundcond}
}
where %and 
the reference frame is chosen such that $\left\langle v(0)\right\rangle_s = 0$, with symmetry $\left\langle v(-z)\right\rangle_s = - \left\langle v(z)\right\rangle_s$. 
Due to translational invariance along $x$ and $y$, Eq.~\eqref{langevinvst} in a steady state yields 
\al{
- \partial_z \,\big\langle \txz(z) \big\rangle_s+ \eta_0\, \partial_z^2 \,\big\langle v(z)\big\rangle_s = 0.
\label{forcebal1}
}
Note that Eqs.~\eqref{langevinphi} and \eqref{langevinvst} introduce a correlation between the fluctuations of $\phi$ and those of $\bm v$. 
The fluctuations of $\bm v$ are the subject of fluctuating hydrodynamics (see, e.g., Ref.~\cite{Dunweg09}), and follow $\ave{(v - \ave{v}_s)^2}_s \sim k_B T/(2\rho {d}^3)$, where ${d}$ is the considered length scale. It is thus an accepted observation that the velocity fluctuations become increasingly irrelevant upon coarse-graining, relative to macroscopic driving such as shear. As stated in Section \ref{dimensionalsect}, this may be seen also upon coarse-graining Eqs.~\eqref{langevinphi} and \eqref{langevinv}, including external shear. The remnant of the fluctuations of $\bm v$ upon coarse-graining are --- via the fluctuation dissipation theorem --- hydrodynamic interactions, e.g., acting between the colloidal particles mentioned above \cite{dhont}. Neglecting these, we thus omit in the following the first term in Eq.~\eqref{langevinv} (the time derivative) as well as the last term due to the noise.

Accordingly, the stress tensor $T^\phi$, a functional $\txz(z)[\bm v]$ of $\bm v$ (via Eq.~\eqref{langevinphi}),  can be found from treating 
$\bm v$ as a given input to Eq.~\eqref{langevinphi}, 
\begin{align}\label{eq:av}
 \big\langle \txz(z) \big\rangle_s= \big\langle \txz(z)[\bm v] \big\rangle_s=\big\langle \txz(z)\big\rangle_s[\langle v \rangle_s],
\end{align}
where the square brackets indicate a functional dependence: the stress tensor at position $z$ depends on the velocity profile. We will in the following omit the use of $\langle \cdots \rangle_s$ when referring to the velocity field for ease of notation. 
Also note that the transverse projection indicated by the subscript $\perp$ in Eq.~\eref{langevinvst} is unnecessary for discussing the mean shear velocity profile.

In the remaining part of this section, we aim at determining $\big\langle \txz(z)\big\rangle_s$ for a  given velocity field. We start with a brief review of linear response theory for a Fokker-Planck equation in Subsec.~\ref{LR:FPeqs}. In Subsec.~\ref{LR:micro}, as a reference, we apply it to the case of Brownian particles. Then, in Subsec.~\ref{LR:FT} we adapt this formalism to the case of the fluctuating fields as in Eq.~\eqref{langevinphi}.

\subsection{Linear response for a Fokker-Planck equation}
\label{LR:FPeqs}

Consider a generic system characterized by phase-space variables $\Gamma$ (below, $\Gamma$ will stand for the positions of Brownian particles, or for the values of the field $\phi(\bm r)$), and  a time dependent probability distribution $\mathcal P(\Gamma,t)$.  This probability obeys a Fokker-Planck equation
\al{
\partial_t \mathcal P = L(\Gamma,t) \mathcal P,
\label{dPdtgeneral}
}
with a Fokker-Planck operator $L(\Gamma,t) = L_e(\Gamma) + \delta L(\Gamma,t)$, containing an equilibrium  part $L_e$ and a perturbation $\delta L$, which may be time-dependent. In the absence of perturbations {(i.e., with $\delta L=0$)}, the equilibrium distribution $\mathcal P_e$ obeys $L_e \mathcal P_e = 0$, and the time-dependent correlation function of two phase space observables $f(\Gamma)$ and $g(\Gamma)$ can be written (see, e.g.,  Ref.~ \cite{risken} for the derivation),
\al{
\ave{f(t)g(0)}_e = \int d\Gamma \;f(\Gamma)e^{L_e t}g(\Gamma)\mathcal P_e(\Gamma).
\label{corrEq}
}
Here, $\langle\ldots\rangle_e$ indicates the average over the equilibrium distribution. Adding a time independent perturbation $\delta L$, switched on at $t=0$, the solution of  Eq.~\eqref{dPdtgeneral} is $\mathcal P(\Gamma,t) = e^{L t} \mathcal P_e$, which can rewritten  \cite{fuchs} by using $e^{L t} = 1 + \int_{0}^t dt'\;e^{L t'}L$ (where we obtain the steady state $\mathcal P_s(\Gamma)$ by letting $t\to\infty$, and using $L_e \mathcal P_e = 0$),
\al{
\mathcal P_s(\Gamma) = \mathcal P_e + \int_0^\infty dt'e^{L t'}  \delta L \,\mathcal P_e.
\label{eq:ev-nlinear}
}
Equation \eqref{eq:ev-nlinear} contains the perturbation to any order, and we obtain the linear response, valid for weak perturbations, by approximating $L$ with $L_e$ in the exponential,
\al{
\mathcal P_s(\Gamma) = \mathcal P_e + \int_0^\infty dt'e^{L_e t'}  \delta L \,\mathcal P_e.
\label{eq:ev-linear}
}
Accordingly, the
steady-state expectation value
\al{
\ave {\Delta f}_{s} &= \int d\Gamma\;\Delta f(\Gamma)\,\mathcal P_s
\label{nessave1}
}
of the induced change $\Delta f = f - \langle f \rangle_e$ of observable $f(\Gamma)$ compared to its equilibrium expectation value
$\langle f \rangle_e = \int d\Gamma\,f(\Gamma)\mathcal P_e$, is straightforwardly written,
\al{
\ave {\Delta f}_{s} &= \int_0^\infty dt'  \int d\Gamma\; f(\Gamma)\, e^{L_e t'}\delta L \,\mathcal P_e.
\label{eq:deltaf-added}
}
The linear response is thus determined by the action of the perturbation $\delta L$ on the equilibrium distribution $\mathcal P_e$. Often (e.g., in the case of shear as considered below, or for potential perturbations  \cite{kubobook})
the equilibrium distribution is an eigenfunction of $\delta L$, i.e., $\delta L \,\mathcal P_e = g(\Gamma) \,\mathcal P_e$, and Eq.~\eref{eq:deltaf-added} reduces to the time-integral of an equilibrium correlation function {(see also Eq.~\eqref{corrEq})},
\al{
\ave {\Delta f}_{s} &= \int_0^\infty dt \; \ave{\Delta f(t)g(0)}_e \nl
&=\int_0^\infty dt \; \ave{\Delta f(t)\Delta g(0)}_e,
\label{nessave2}
}
{where the last equality follows as $\ave{\Delta f(t)}_e = 0$, and this relation eventually involves only the fluctuations of $f$ and $g$}.

\subsection{Viscosity from a microscopic model}
\label{LR:micro}

As a reference, we first apply the linear response theory recalled in the previous section to  $N$ over-damped Brownian particles located at the positions $\bm r_1, \ldots \bm r_N$ in space, i.e., $\Gamma = \{\bm r_i\}$, $i = 1,\ldots,N$.  
Here, the equilibrium Fokker-Planck operator $L(\Gamma,t)_e$ reads  \cite{risken} 
\al{
L_e&= D_0 \sum_{i=1}^N \bm \nabla_i \cdot (\bm \nabla_i - \beta \bm F_i),
\label{eq:Omega-e}
}
where $D_0$ is the bare diffusion coefficient and $\bm F_i = (F_{ix},F_{iy},F_{iz})$ indicates the force acting on particle $i$. The force is given by $\bm F_i = - \bm \nabla_i U(\Gamma)$, where the potential $U$ results from both the interparticle interaction and the possible external potentials. (The equilibrium distribution is accordingly given by  $\mathcal P_e\propto e^{-\beta U}$.) The presence of an imposed velocity field $\bm v(\bm r)$ of the solvent  gives rise to the perturbation  \cite{risken,dhont},
\al{
\delta L&= -\sum_{i=1}^N \bm \nabla_i \cdot \bm v({\bm r}_i).
}
For a system under shear with $\bm v = v(z)\hat x$ (as in Eq.~\eqref{vxdef}), the linearized steady state distribution is determined by Eq.~\eref{eq:ev-linear}, i.e., (here we used $\delta L \mathcal P_e= -\sum_i^N v(z_i) F_{ix}\mathcal P_e$),
\al{
\mathcal P_s(\Gamma) = \mathcal P_e - \beta \sum_{i=1}^N \int_0^\infty dt'e^{L_e t'}  v(z_i) F_{ix} \mathcal P_e.
\label{eq:ev-linearRENAMEME}
}
The induced change of  observable $f$ in the steady state, due to the presence of the fluid flow, is therefore, up to linear order, given by Eq.~\eqref{nessave2},  
\al{
\ave {\Delta f}_{s} = -\beta\int_0^\infty dt'\;  \left\langle \Delta f(t') \sum_{i=1}^N v(z_i) F_{ix} \right\rangle_{e}.
\label{nessave1RENAMEME}
}
In the case of homogeneous shear \cite{fuchs}, with $v(z) = s_0 z$ and constant shear rate $s_0$, the above relation can be written in terms of the potential part (omitting kinetic contributions in the overdamped regime) of the microscopic stress tensor, $T_{xz}^{(s_0)}=-\sum_i z_i F_{ix}$. In the present case of inhomogeneous shear with a position-dependent shear rate $s(z)$,  it is natural to define a local stress tensor $T_{xz}$ involving only particles at height $z$, i.e.,
\al{
T_{xz}(z) =  -\sum_{i=1}^N \delta(z-z_i) z_i F_{ix}.
}
In terms of this,
\al{
\ave {\Delta f}_s&= \beta \int_0^\infty dt\int dz' \;  \frac{v(z')}{z'}\left\langle \Delta f(t) T_{xz}(z',0) \right\rangle_e.
\label{NESSave}
}
In particular, since $\ave{T_{xz}( z)}_e = 0$, 
which follows directly from spatial symmetries,  
the steady state expectation value of the local stress tensor at height $z$ is 
\al{
\ave{T_{xz}( z)}_s&= \beta\int_0^\infty dt\int dz' \;  \frac{v(z')}{z'} \left\langle T_{xz}( z,t)\, T_{xz}(z',0)\right\rangle_e,
\label{nessavestress}
}
up to the linear order in the perturbation. Note that the integrand on the r.h.s.~of this equation involves the two-time correlation function of the stress tensor. 
Equation \eqref{nessavestress} is a generalisation of the well-known Green-Kubo relation to {the case of} systems with spatially dependent shear rate $s(z)$ (see Eq.~\eqref{eq:def-s}). It allows one to express the %for 
shear viscosity $\eta$ {in terms of correlation functions. In fact,}  for a constant $s(z) = s_0$, on the basis of the definition in Eq.~\eqref{eta}, we recover the well-known result \cite{kubobook,fuchs}
\al{\label{eq:eb}
\eta = \frac{\ave{ T_{xz}}_s}{s_0} = \beta\int_0^\infty dt\;  \left\langle T_{xz}(t)\, T_{xz}(0)\right\rangle_e.
}
In the bulk, this is indeed the definition of viscosity, which corresponds exactly to our effective viscosity $\etaeff$ 
in Eq.~\eqref{eta}. 
This is because the force per area acting on the upper plate is given by $\ave{ T_{xz}(z=L/2)}_s$, so that for a bulk, Eq.~\eqref{eta} becomes equal to Eq.~\eqref{eq:eb}.

\subsection{Viscosity of fluctuating fields}
\label{LR:FT}

Let us now consider Eqs.~\eref{langevinphi} and \eref{modAB2} for the dynamics of {the field} $\phi$ (coupled to $\bm v$). The advection term in Eq.~\eref{langevinphi} with  an assigned velocity profile $\bm v$ is treated as a small perturbation to the equilibrium dynamics of $\phi$.

We first find the Fokker-Planck equation for the probability functional $\mathcal P[\phi]$, associated with Eqs.~\eqref{langevinphi} and \eqref{modAB2}. It takes the form
\al{
\partial_t \mathcal P[\phi] &= \hat L^\phi \mathcal P[\phi] = \left(\hat L^\phi_e + \delta \hat L^\phi%(t)
\right) \mathcal P[\phi], 
}
where the Fokker-Planck operator $\hat L^\phi$ is naturally split into a term $\hat L_e^\phi$ corresponding to the dynamics in the absence of advection (see Appendix \ref{potcondsect} for details, in particular Eq.~\eqref{diffterms}), 
i.e., 
\al{\hat L_e^\phi\mathcal  P &=  \int d\bm r \frac{\delta}{\delta \phi} \hat \mu \left[ \frac 1 \beta \frac{\delta}{\delta \phi}+ \frac{\delta}{\delta \phi} \mathcal H \right] \mathcal P,\cm
}
and a contribution due to advection (see Eq.~\eqref{straemterms}) related to the non-equilibrium perturbation
\al{
\delta \hat L^\phi \mathcal P &= \int d\bm r \; \left[{\bm \nabla} \cdot (\bm v\phi) \right] \frac{\delta}{\delta \phi} \mathcal P .
}

Here, the Boltzmann distribution $\mathcal P_e[\phi] \propto e^{-\beta \mathcal{H}[\phi]}$ associated with the Hamiltonian in Eq.~\eref{H1} is the equilibrium distribution of the system with $\bm v = \bm 0$,  and {therefore it satisfies the condition}  
$\hat L^\phi_e \mathcal P_e[\phi] = 0$.

Switching on the perturbation due to $\bm v$ at $t=0$, the resulting distribution $\mathcal P[\phi](t)$ at a later time is formally determined by
$\mathcal P[\phi](t) = e^{t\hat L^\phi }\mathcal P_e[\phi]$
and a linearization of the evolution operator in the perturbation $\bm v$ renders an expression analogous to Eq.~\eqref{eq:ev-linear},
\al{
\mathcal P[\phi](t) = \mathcal P_e[\phi] +\int_0^t dt'\;e^{ t' \hat L_e^\phi } \delta \hat L^\phi\mathcal P_e[\phi] + \mathcal{O}(v^2).
\label{pphit}
}
The perturbation acts on $\mathcal P_e$ as (see Appendix \ref{potcondsect})
\al{
\delta \hat L^\phi \mathcal P_e[\phi] &= - \beta \int d\bm r \; \left[\bm\nabla \cdot  (\bm v  \phi)\right]  \frac{\delta \mathcal H}{\delta \phi} \mathcal P_e[\phi]\nl
&=\beta \int d\bm r \; \bm v \cdot \left[ \phi  \bm\nabla \frac{\delta \mathcal H}{\delta \phi} \right]\mathcal  P_e[\phi].
}

We identify the stress tensor in Eq.~\eqref{fvequalsTphi}, $\phi  \bm\nabla (\delta \mathcal H/\delta \phi) = \bm \nabla \cdot T^\phi$. 
Since $\bm v \cdot ( \bm\nabla \cdot T^\phi ) = \sum_{ik} v_i \big(\partial_k T^\phi_{ki}\big)$ and $\bm \nabla \cdot \bm v = 0$, integration by parts gives
\al{
\delta \hat L  \mathcal P_e 
&= -\beta \left[\sum_{ik} \int d\bm r \; \big(\partial_k v_i \big) T^\phi_{ki}\right] \mathcal  P_e \nl
&= - \beta\left[ \int d\bm r \; \big( \bm \nabla \bm v \big) : T^\phi\right] \mathcal  P_e .
}
As before, this allows us to calculate  
the average of an observable $f$  
in the steady state ($t\to\infty$),
\al{
&\ave{\Delta f({\bm r})}_{s} =  \nl
&- \beta\int_0^\infty \!\!\!dt \int \!\!d\bm r' \; ({ \bm \nabla_{\bm r'} \bm v(\bm r') }) : \ave{   \Delta f({\bm r}, t)T^\phi(\bm r',0 )}_{e} ,
\label{r1}
}
which is analogous to Eq.~\eqref{NESSave}.
{Equation~\eqref{r1} generalizes to the case of a fluctuating field. A similar relationship has been derived  in the context of Liouville dynamics \cite{procaccia1979,procaccia1980,onukibook} for microscopic systems subject to slowly varying velocity gradients.}
In the case of a shear velocity profile such as the one in Eq.~\eref{vxdef}, the previous equation %this 
becomes 
\al{
\ave{\Delta f({\bm r})}_{s}  = -\beta \int d\bm r' { s(z')}\int_0^\infty dt \,\ave{ \Delta f({\bm r},t) T^\phi_{xz} (\bm r',0) }_e.
\label{generallr}
}
The local shear stress required in Eq.~\eqref{forcebal1} finally follows by replacing $f$ by $T_{xz}^\phi$ (we recall that 
${\ave{ T_{xz}^\phi(z)}_{e}} =0$),
\al{
&{\ave{ T_{xz}^\phi(z)}_{s}} =  - \beta 
\int d \bm r _\parallel '  \int_{\D} dz'   \nl
&\qquad \qquad \int_0^\infty dt \,s(z')\ave{ T_{xz}^\phi(\bm r,t) T_{xz}^\phi(\bm r',0) }_e,
\label{linresp}
}
{where $\bm r = (\bm r_{\|},z)$ and analogous decomposition for $\bm r'$.}
Note that on the l.h.s.~we have omitted the dependence on $\bm r_\parallel$, as it disappears in the steady state due to its translational invariance --- see Sec.~\ref{fsect}. Analogously, the correlation function on the r.h.s.~depends  on $\bm r_\parallel-\bm r_\parallel'$, $z$ and $z'$.
Accordingly, in order to determine the expectation value of the stress tensor in the non-equilibrium {stationary state} in the presence of a weak fluid flow --- which we need in order to compute the effective viscosity according to Eq.~\eqref{forcebal1} --- we calculate below correlation functions of the stress tensor in {equilibrium}. 
Note that in Eq.~\eqref{linresp}, the local shear rate $s(z')$ appears, while in the analogous expression in Eq.~\eqref{nessavestress} for the set of Brownian particles, the microscopic counterpart $v(z')/z'$ shows up.  
{Note also that, compared to Eq.~\eref{nessavestress}, Eq.~\eref{linresp} involves an additional integral over $\bm r_\parallel$ which comes from the Fokker-Planck operator (see Appendix \ref{potcondsect}).}

\subsection{Non-local Stokes equation}
\label{stokessect}

As an interesting observation, we note that Eq.~\eqref{langevinv} in the stationary state (or, equivalently, Eq.~\eqref{forcebal1}) for the geometrical setting of Fig.~\ref{fig:confinedshear} may be cast in the form
\al{
0 = \frac{d}{dz}  \int_\D d z' \eta(z,z') \frac{d}{dz'} v(z'),  
\label{nonlinstokes1}
}
where a non-local viscosity kernel $\eta(z,z')$ appears, once the linear response relation in Eq.~\eqref{linresp} has been used together with the definition in Eq.~\eqref{eq:def-s}. This equation may be viewed as a non-local Stokes equation; it is a consequence of the fact that Eq.~\eref{langevinv} for $\bm v$ has the form of a continuity equation. In turn, Eq.~\eref{nonlinstokes1} amounts to requiring that the shear stress $\int_\D d z' \eta(z,z') s(z')$ is  independent of the coordinate $z$, i.e., that it is constant across the {film}. %system.
In particular, $\eta(z,z')$ takes the explicit form  
\al{
\eta(z,z') &= \delta(z-z')\eta_0 \notag\\&+ \beta %\frac {} {k_B T} 
\int_0^\infty dt\int d \bm r _\parallel '\ave{  T_{xz}^\phi(z,t)  T_{xz}^\phi(z',0) }_e,
\label{nonlinstokes2}
}
which shows that the local contribution $\eta_0$ arising from the pure solvent is effectively modified by a {potentially} {non-local} correction due to the coupling %$\propto \lambda$ 
to the velocity field in Eq.~\eqref{langevinphi}  and to the correlations of the field $\phi$.
{In fact, $\eta(z,z')$ in Eq.~\eqref{nonlinstokes2} does not necessarily vanish for $z\neq z'$.}
Accordingly, correlated fluctuations of the order parameter {are expected to cause} %induce 
non-local effects in the hydrodynamics of the solvent. 

\section{Effective viscosity of Model \ha in confined geometry}
\label{resultssect}

In this section 
we apply the linear response formalism derived in Sec.~\ref{linrespsect} to Model \ha {within the Gaussian approximation, as discussed in Sec.~\ref{coarsegrainsect}}. This leads to a set of self-consistent equations for the shear rate, as set out in Subsec.~\ref{selfconsistsect}. We discuss the exact autocorrelation functions which arise in the model in Subsec.~\ref{modelsect} and present our {predictions} 
for the dependence of the effective viscosity $\etaeff$ on the correlation length $\xi$ in Subsec.~\ref{corrsect}.

\subsection{Self-consistent equation for the shear rate}
\label{selfconsistsect}

The condition of homogeneous, $z$-independent stress imposed by Eq.~\eqref{forcebal1}  (or, equivalently, Eq.~\eqref{nonlinstokes1} after having used the linear response relation in Eq.~\eqref{linresp}) provides a self-consistent equation for the shear rate $s(z)$ in Eq.~\eqref{eq:def-s}, after integration of the equation along $z$ (we already point to the dependence on correlation length $\xi$), 
\al{
s(z) + \frac {1} {\eta_0} \beta \int_\D dz' \;\; {s(z')}{\czzp} = \mbox{const.},
\label{sselfconsist}
}
where we introduced
\al{
\czzp = \int_0^\infty dt\int d \bm r _\parallel ' \;\ave{  T_{xz}^\phi(\bm r,t) T_{xz}^\phi(\bm r',0) }_e.
\label{czzpdef}
}
Equations \eref{sselfconsist} and \eref{czzpdef} can be solved self-consistently subject to the boundary conditions for $v$ and $\phi$ at the confining surfaces of the film. 
Note that the constant in Eq.~\eqref{sselfconsist} can be readily identified with $T_{tot}/\eta_0$, where $T_{tot}$ is the total stress. It contains the part due to shearing the solvent with viscosity $\eta_0$, and the stress arising from the order parameter field.   
In terms of it, the effective viscosity $\eta_{\rm eff}$ defined in Eq.~\eqref{eta} is {eventually} %finally 
given by
\begin{align}
\eta_{\rm eff}=\frac{T_{tot}}{s_0}.
\label{eq:eta-eff}
\end{align}
We now discuss the specific model and solve Eq.~\eqref{sselfconsist} for that case. 

\subsection{Model and stress-tensor autocorrelation function}
\label{modelsect}

As anticipated in Sec.~\ref{sec:intro}, our aim is to investigate how confined and correlated fluctuations affect the effective viscosity. We consider a simple model within which correlations occur,  namely Model \ha discussed in Sec.~\ref{sec:modH} with a Gaussian effective Hamiltonian corresponding to Eq.~\eqref{H1b} with $u=0$, i.e.,
\al{
\mathcal H [\phi]= \frac{1}{2} \int d\bm r \; \left[ (\nabla \phi)^2 + \xi^{-2} \phi^2(\bm r) \right].
\label{Hgauss}
}
As noted above, we use Dirichlet boundary conditions for $\phi$ at the two walls in Fig.~\ref{fig:confinedshear}, i.e., $\phi(z=\pm L/2)=0$ and the stick boundary conditions for $v$ given by Eq.~\eqref{vboundcond}.

In order to calculate $\ave{T_{xz}^\phi(z)}_{s}$, the linear response relation in Eq.~\eqref{linresp} requires the knowledge of the correlation function of the $xz$-component of the field stress tensor from Sec.~\ref{fsect}, which, {for $\cal H$ in Eq.~\eqref{Hgauss},} 
is given by
\al{
\txz (\bm r) = \partial_x \phi\, \partial_z \phi. 
}
The equilibrium stress-tensor autocorrelation function $\czzp$ in Eq.~\eqref{czzpdef} required for the calculation of the total stress $T_{tot}$ from Eq.~\eref{sselfconsist} thus comprises suitable derivatives of a four-point correlation function of the order parameter field $\phi$. {For a Gaussian Hamiltonian such as \eqref{Hgauss}, the latter can} 
be calculated by 
using Wick's theorem, as detailed in Appendix \ref{stresstenappend}. In turn, this requires the knowledge of the two-point and two-time correlation function {$\langle \phi(\bm r,t)\phi(\bm r',t')\rangle_e$} of the order parameter in equilibrium within the film of thickness $L$, with Dirichlet boundary conditions at the confining surfaces. 

{In order to determine the latter, it is convenient to carry out the analysis} 
in Fourier space, where $\bm p$ is the spatial wave vector conjugate to $\bm r_\parallel = (x,y)$, and the frequency $\omega$ results from the Fourier transform in time (which can be introduced because equilibrium correlation functions are time-translational invariant). 
The {integral $C$ of the} stress correlation function defined in Eq.~\eqref{czzpdef} turns out to be given by
(see  Appendix \ref{FTappend}) 
\al{
\czzpl& = \frac{( k_B T)^2}{\mu(2\pi )^2} \frac 1{L^2} \int_{0}^{\Lambda L} d\tilde p \int d\tilde \omega\;\; \tilde f(\tilde z,\tilde z', \tilde \omega,\tilde p, \tilde \xi)\nl
&= \frac{( k_B T)^2}{\mu(2\pi )^2} \frac 1{L^2} {\czzplr},
\label{czzpfilmr}
}

where $\czzplr$ is a dimensionless function and $\mu$ is the mobility appearing in the Langevin equation of Model H(A); see the discussion after Eq.~\eqref{modAB}. We have introduced 
a cutoff $\Lambda = 1/a$, set by the microscopic length scale $a$, and have rescaled to dimensionless variables 
\al{\tilde \omega = L \omega/\mu, \quad \tilde p = \l0 p,\quad \tilde z =  z/L,} and similarly for the remaining length scales, i.e., $\tilde z' = z'/L$, $\tilde \xi = \xi / L$, {and $\tilde \Lambda = \Lambda L$. Additional details as well as the explicit expression of} 
the integrand $\tilde f$ are provided in Appendix \ref{rescaleappend}.

The integrals in Eq.~\eqref{czzpfilmr} can now be computed numerically as functions of $\tilde z$, $\tilde z' \in \tilde\D=[-1/2,1/2]$ for various values of  the dimensionless cutoff $\tilde \Lambda$, indicated by the superscript. 
The function $\czzpl$, which essentially determines the non-local contribution to the viscosity in Eq.~\eqref{nonlinstokes1}, %, $\eta(z,z') = \frac {\lambda} {k_B T} \czzpl$, 
is indeed non-local for finite $\Lambda$ (and therefore $\tilde\Lambda$). 
However, upon increasing $\tilde \Lambda$, $\czzplr$ becomes increasingly local in that at $\tilde z = \tilde z'$ the  $\tilde p$-integral in Eq.~\eqref{czzpfilmr} turns out to be ultraviolet divergent, as shown in Appedix \ref{exampleappend} (see, in particular, Fig.~\ref{fig:czzpLam} therein). 
A consequence of this divergence is that integrals 
{involving $\czzplr$ can be expressed in terms of an envelope function for asymptotically large $\tilde \Lambda$.}
{In particular, for}
%For 
a smooth function $\psi(\tilde z)$, we find for $\tilde\Lambda \gg 1$,
\al{
\int_{-1/2}^{1/2} d\tilde z' \;\psi(\tilde z')\czzplr \;\stackrel{\tilde\Lambda \gg 1}{\simeq} \;g^{\tilde \Lambda}(\tilde z, \tilde \xi) \psi(\tilde z),
\label{ansatz}
}
where 
\al{
g^{\tilde \Lambda}(\tilde z, \tilde \xi) &= \int_{-1/2}^{1/2} d\tilde z' \;\czzplr.
}
The envelope function $g^{\tilde\Lambda}(\tilde z, \tilde \xi)$ takes two different values depending on whether $\tilde z$ is at the boundaries $\tilde z = \pm 1/2$ or not, with the latter value scaling linearly with the cutoff $\tilde \Lambda$,
\al{
g^{\tilde \Lambda}(\tilde z, \tilde \xi) &= \int_{-1/2}^{1/2} d\tilde z' \;\czzplr \nl
&\xrightarrow[]{\tilde \Lambda \gg 1} \; 
\begin{cases}
\simeq 0.39\;\tilde \Lambda &\mbox{for}\quad \tilde z \in (-1/2,1/2),\\
0 &\mbox{for}\quad \tilde z =\pm 1/2.
\end{cases}
\label{gcases}
}
{
Eqs.~\eref{ansatz} and \eref{gcases} hold as long as the correlation length is large compared to the microscopic cutoff, i.e., $\tilde \xi \gg \tilde \Lambda^{-1}$. Hence we drop the dependence of $g^{\tilde \Lambda}$ on $\tilde \xi$.
}
(Note that the limit $\tilde\Lambda \gg 1$ actually corresponds to considering the fluid confined within a film of large thickness $L \gg a$.) Indeed, this suggests that
{$\czzplr  \to g^{\tilde\Lambda}(\tilde z,\tilde \xi)\delta(\tilde z'-\tilde z)$ upon increasing $\tilde \Lambda$.}

A plot of $g^{\tilde \Lambda}(\tilde z,\tilde \xi = \infty)$  for various values of $\tilde \Lambda$ is reported in Fig.~\ref{fig:g(z)} of Appendix \ref{exampleappend}, where we also discuss in detail Eq.~\eqref{ansatz}.
We conclude that, in our model, bulk effects dominate the computed viscosity kernel $\eta(z,z')$. Next, we analyze the resulting velocity profile. 

Using Eq.~\eref{ansatz}, the self-consistent equation \eref{sselfconsist} has a simple algebraic solution $s^{\tilde\Lambda}(z)$ {for $\tilde\Lambda \gg 1$}, %in the limit $\tilde \Lambda\to \infty$, 
\al{
s^{\tilde \Lambda} ( z) & = \frac{T_{tot}/\eta_0}{1 + \alpha g^{\tilde \Lambda}(\tilde z)}.
\label{eq:sr-largeL}
}
For convenience we introduced here
\al{
\alpha = \frac{1}{(2\pi)^2\beta \mu\eta_0 L},
\label{eq:defalpha}
}
which is a dimensionless parameter arising from rescaling, noise correlators and prefactors in the linear response relation. 
{In fact, from the Langevin equations \eqref{langevinphi} and \eqref{langevinv}, combined with Eq.~\eqref{Hgauss}, one infers that $\mu$ has dimensions of (length)$^2$/time, $\eta_0$ of (energy $\times$ time)/(length)$^3$, and therefore 
\al{
\ell \equiv \frac{1}{\beta\mu\eta_0}
\label{eq:def-ell}
}
is  a length scale which makes $\alpha \propto \ell/L$ in Eq.~\eqref{eq:defalpha} dimensionless.} 
$\ell$ is related to the mobility of the order parameter field, and can be seen as the effective hydrodynamic radius with respect to $\eta_0$. 
The total stress $\tt$ (yet undetermined by this analysis) is fixed by imposing 
{the velocity $v(z)$ resulting from the integration of the shear rate $s(z)$ in Eq.~\eqref{eq:sr-largeL} (see Eq.~\eqref{eq:def-s}) to be equal to $\pm v^*$ at the boundaries (Eq.~\eqref{vboundcond}), which implies}
\al{
\frac{\tt}{\eta_0} =\frac{v^*} L \left[  \int_0^{1/2} d\tilde z\,  \frac{1}{1 + \alpha g^{\tilde \Lambda}(\tilde z)} \right]^{-1}.
\label{Tscaling}
}
{For $\tilde \Lambda\gg 1$,}
$\tt\sim\alpha\tilde\Lambda$
due to Eq.~\eref{gcases}, while Eq.~\eqref{eq:sr-largeL} implies that a linear velocity profile is recovered with shear rate
\al{
s^{\tilde \Lambda}( z)  \xrightarrow[]{\tilde \Lambda\gg 1} 2v^*/L = s_0.
\label{slimit}
}
Here $s_0$ is the shear rate for a linear velocity profile with velocities $\pm v^*$ at the surfaces; see Sec.~\ref{syssect}. This is a consequence of the fact that $\eta(z,z')$ is eventually dominated by (local) bulk contributions. 
{The linear increase of $T_{tot}$ as a function of $\tilde\Lambda \gg 1$ carries over to the effective viscosity determined from Eq.~\eqref{eq:eta-eff}, which 
formally diverges. As stated, this holds also for finite correlation lengths, as long as $\xi \gg \Lambda^{-1} $.}

The above analysis of model \ha under confinement reveals a dependence of the effective viscosity $\eta_{\rm eff}$  on the 
cutoff $\Lambda$ for macroscopic correlation lengths $\xi$. Within this simple 
model, {introducing this cutoff is just a convenient way} 
of taking into account the microscopic structure of the system under study. 
We expand on the cut-off dependence in  Appendix \ref{critappend}. 
The physical interpretation of this strong dependence of $\eta_{\rm eff}$ on $\Lambda$ is that the contributions of fluctuations to this quantity are essentially determined by the material properties of the system under investigation, which the simplified mesoscopic description investigated here is unable to capture in detail.

However, $\eta_{\rm eff}$  might still display a dependence on the correlation length $\xi$ and eventually on the thickness $L$ of the film within which fluctuations are constrained. In other words, universal (and non-local) aspects  may appear in the interplay between $L$ and $\xi$, and can be extracted by considering the dependence of the effective viscosity on the correlation length $\xi$, as we do further below.

\subsection{Dependence of the effective viscosity on the correlation length}
\label{corrsect}

\begin{figure}[t]
 \begin{center}
 \includegraphics[width=\columnwidth]{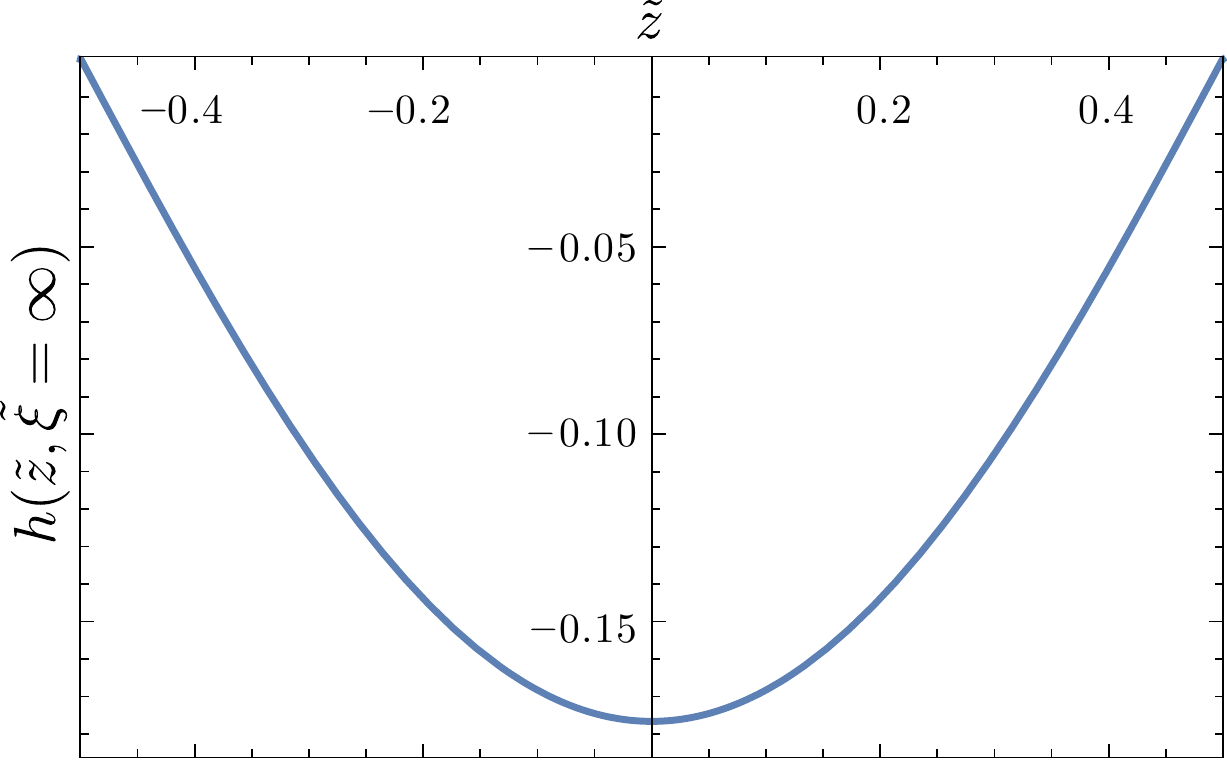}
  \caption{Dependence of the function $h(\tilde z)$ (see Eq.~\eqref{h}) on the reduced coordinate $\tilde z$ across the film (with $|\tilde z|\leq 1/2$) at criticality, i.e., for $\tilde \xi = \infty$. $h(\tilde z)$ is related to the derivative of the non-local viscosity $\eta(z,z')$ with respect to correlation length $\xi$, (see, e.g., Fig.~\ref{fig:czzpLam}), after integration over $\tilde z'$ (up to prefactors of $\tilde \xi$) and it turns out to become independent of the cutoff $\tilde \Lambda$ for $\tilde\Lambda \gg 1$.} 
 \label{fig:h}
 \end{center}
\end{figure}

In order to distill the effects of large-scale fluctuations, we differentiate
Eq.~\eref{Tscaling} with respect to (the rescaled) correlation length $\tilde \xi = \xi / L$,
\al{
\frac 1 {s_0\eta_0} \partial_{\tilde \xi} T_{tot} =&\alpha \frac{\int_{0}^{1/2} d\tilde z \;   \left[1+\alpha g^{\tilde \Lambda}(\tilde z)\right]^{-2} \partial_{\tilde \xi} g^{\tilde \Lambda}(\tilde z)}{\left( \int_{0}^{1/2} d\tilde z \; \left[ 1+\alpha g^{\tilde \Lambda}(\tilde z)\right]^{-1}\right)^2} .
\label{dxiTcrit}
}
Since $g^{\tilde \Lambda}(\tilde z)$ become spatially constant for large cutoffs --- see Eq.~\eref{ansatz} ---  we are left with computing 
\al{
\partial_{\tilde \xi} g^{\tilde \Lambda}(\tilde z) = \int_\D%{-1/2}^{1/2} 
d\tilde z' \;\; \partial_{\tilde \xi}\czzplr \equiv -\frac 1 {{\tilde \xi}^3} h(\tilde z,\tilde \xi).
\label{h}
}
This form arises since $ \partial_{\tilde \xi} \czzplr = \int_0^{\tilde \Lambda} d\tilde p\int d\tilde \omega \; \partial_{\tilde \xi} \tilde f$,
where $\tilde f$ is the same as in Eq.~\eqref{czzpfilmr}. 
(See also Eq.~\eref{Qdef} in Appendix \ref{stresstenappend}.) The derivative of $\tilde f$ with respect to $\tilde\xi$ removes the large-$\tilde p$ divergence of the corresponding integral, mentioned before Eq.~\eqref{ansatz}, and therefore this expression has a well-defined limit as $\tilde\Lambda \to\infty$. In other words, the corresponding functional dependence on $\tilde\Lambda$ is independent of the microscopic details of the system. 
Physically, the function $h(\tilde z,\tilde \xi\,)$  is the integral across the system of the derivative with respect to $\tilde \xi$ of $\eta(z,z')$ in the Stokes equation, and is shown in Fig.~\ref{fig:h} as a function of $\tilde z \in [-1/2,1/2]$ for $\tilde \xi = \infty$. We have therefore identified the contribution in $\eta(z,z')$, which is sensitive to confinement.
{Due to the Dirichlet boundary conditions, $\phi(\tilde z = \pm 1/2) = 0$, it follows that $h(\tilde z = \pm 1/2,\tilde \xi\,) = 0$.} 
%%%%%%%%%%%%%%%%%%%%%%%%%%%%%%%%%%
%%%%%%%%%%%%%%%%%%%%%%%%%%%%%%%%%%

\begin{figure}[t]
 \centering
 \includegraphics[width=\columnwidth]{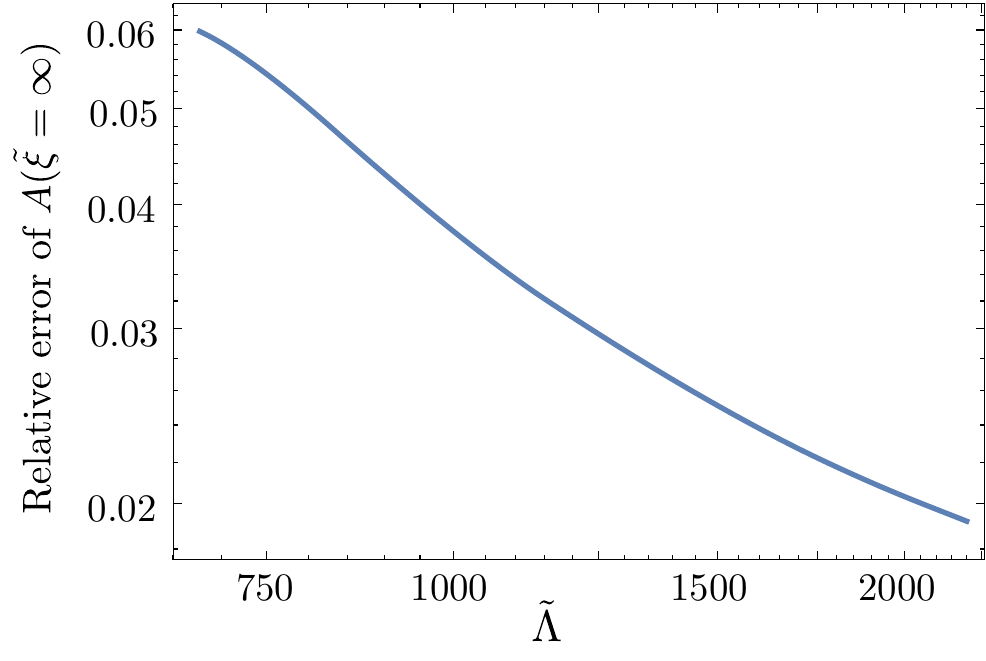}
 \caption{
 Dependence of the function $A(\tilde\xi)$ (see Eq.~\eqref{eq:def-A}), calculated for $\tilde \xi = \infty$, on $\tilde \Lambda$. We show the relative deviation $\big[ 0.113 -A(\tilde\xi=\infty) \big]/0.113$ from the asymptotic numerical value 0.113 in Eq.~\eref{amplitude}, which is approached as $\tilde \Lambda\to\infty$. }
 \label{fig:dxiTtot}
\end{figure}

\begin{figure}[t]
 \begin{center}
 \includegraphics[width=\columnwidth]{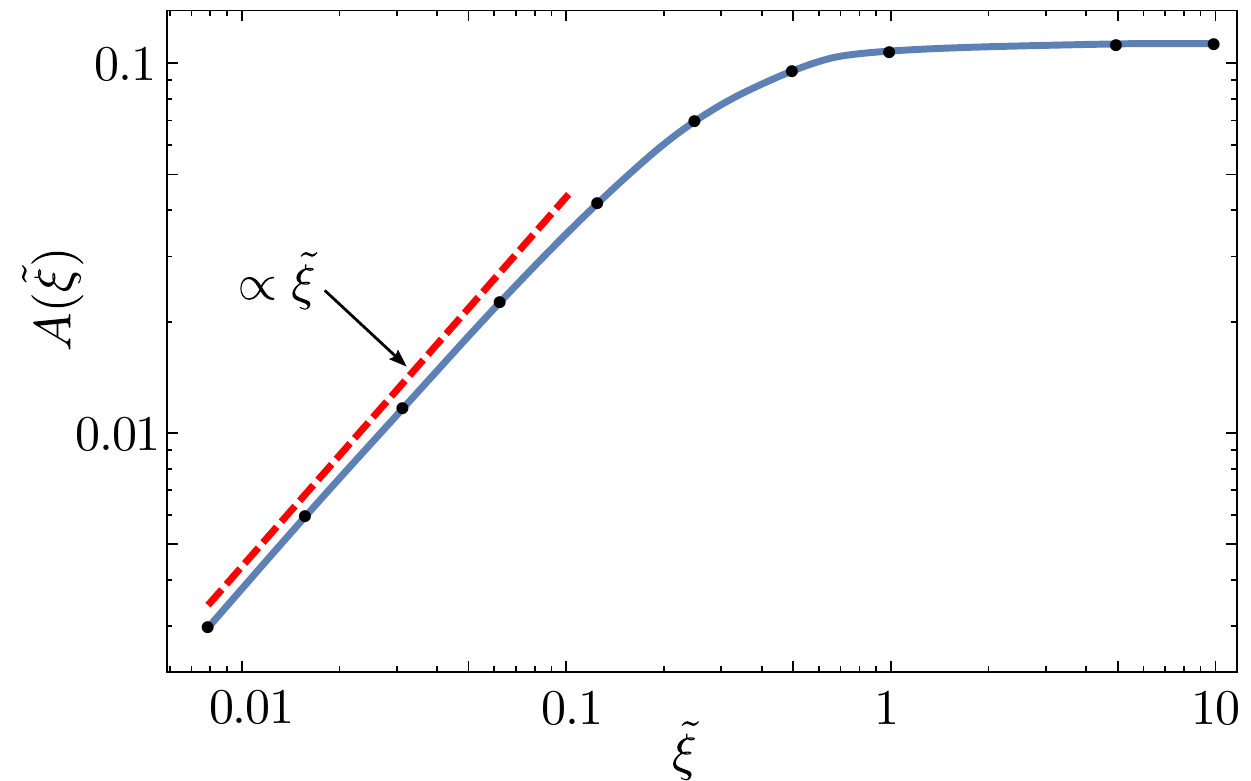}
  \caption{Dependence of the amplitude $A(\tilde \xi)$ on $\tilde\xi$, calculated for $\tilde\Lambda = \infty$. Symbols represent numerical data points, and the blue line is a guide to the eye. The function is linear for $\tilde\xi\ll 1$ (as indicated by the dashed red line), and approaches a finite value for $\tilde\xi\gg 1$. These limits correspond to the case of correlation lengths much shorter and longer than the thickness of the film $L$, respectively. }
\label{fig:A(xi)}
 \end{center}
\end{figure}

\begin{figure}[t]
 \begin{center}
 \includegraphics[width=\columnwidth]{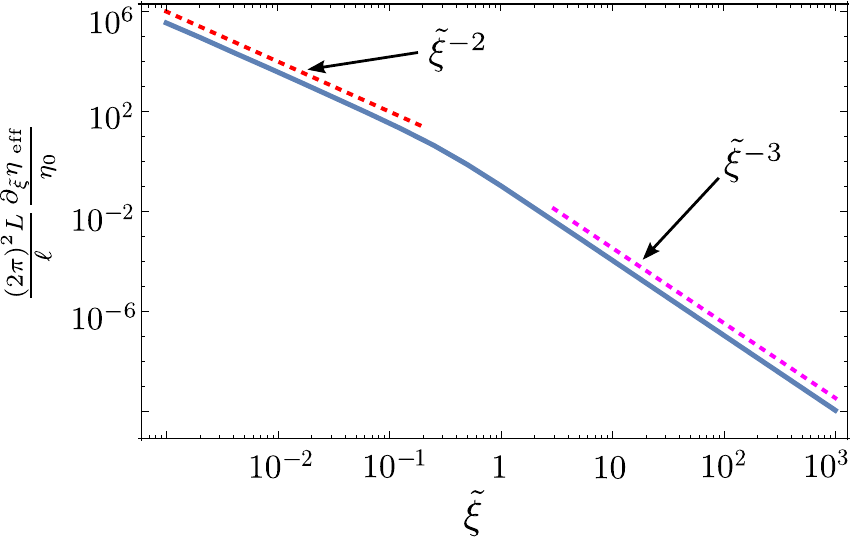}
 \caption{{The derivative of the effective viscosity $\etaeff$ with respect to $\tilde \xi = \xi/L$, as a function of this variable, for $\tilde\Lambda \to\infty$; see Eq.~\eref{ca}. The dashed lines highlight the indicated algebraic behaviours. } 
} %
 \label{fig:dTdxi(xi)}
 \end{center}
\end{figure}

Returning to Eq.~\eref{dxiTcrit}, we obtain in the limit $\tilde \Lambda \to \infty$,
\al{
\frac {\partial_{\tilde \xi} T_{tot}} {s_0\eta_0}  %\xrightarrow[]{\tilde \Lambda\to\infty}&
= -\alpha {\tilde \xi}^{-3} \int_{0}^{1/2} d\tilde z \;h (\tilde z,\tilde \xi\,) = \alpha  \tilde \xi^{-3} A(\tilde \xi).
\label{eq:def-A}
}
For sufficiently large $\tilde \Lambda$, $h(\tilde z,\tilde \xi)$ and therefore the amplitude $A(\tilde \xi)$ is {\it independent} of $\tilde \Lambda$, i.e., %independent 
of the microscopic details of the system. This is shown in Fig.~\ref{fig:dxiTtot} for $\tilde \xi = \infty$, but holds for all values of the correlation length. The dependence of $A$ on $\tilde\xi$, reported in Fig.~\ref{fig:A(xi)}, is characterized by a marked crossover between the following two limiting behaviors:
\al{
A(\tilde \xi) \approx 0.113 \begin{cases}
                    {\tilde \xi}&\mbox{for}\quad \tilde \xi \ll1,\\
                    1&\mbox{for}\quad\tilde \xi \gg1.
                   \end{cases}
\label{amplitude}
}
\noindent Through Eqs.~\eref{eq:eta-eff} and \eqref{eq:def-A}, one now finds the derivative of the effective viscosity, defined in Eq.~\eqref{eta} (not to be confused with the non-local viscosity entering Eq.~\eqref{nonlinstokes1}),
\al{\label{ca}
\frac{\partial_{\tilde \xi} \etaeff}{\eta_0} =  \alpha \tilde \xi^{-3} A(\tilde \xi) = \frac{1}{(2 \pi)^2}\frac{\ell}{L}\times
		   \begin{cases}
                    0.37{\tilde \xi}^{-2}&\mbox{for}\quad \tilde \xi \ll1,\\
                    0.12{\tilde \xi}^{-3}&\mbox{for}\quad \tilde \xi \gg1.
                   \end{cases}
}
This is shown in Fig.~\ref{fig:dTdxi(xi)} from the full numerical solution of Eq.~\eref{dxiTcrit}, where the limits of  Eq.~\eqref{ca} are apparent. 
%
%
%%%%%%%%%%%%%%%%%%%%%%%%%%%%%%%%%%
%%%%%%%%%%%%%%%%%%%%%%%%%%%%%%%%%%
%%%%%%%%%%%%%%%%%%%%%%%%%%%%%%%%%%
%%%%%%%%%%%%%%%%%%%%%%%%%%%%%%%%%%
%
%
Integrating this derivative between two arbitrarily chosen limits, $\tilde \xi_1$ and $\tilde \xi_2$, yields
\al{
&\etaeff(\tilde \xi_2) - \etaeff(\tilde \xi_1) =
\int_{\tilde \xi_1}^{\tilde \xi_2} d\tilde \xi\;\partial_{\tilde \xi} \etaeff  =  \frac{1}{(2 \pi)^2 \beta \mu L} \times \nl
&
		   \begin{cases}
                    0.37\big({\tilde \xi_1}^{-1}- {\tilde \xi_2}^{-1}\big),&\;\;\tilde \xi_1,\tilde \xi_2 \ll1,\\
                    -1.36 + 0.37 {\tilde \xi_1}^{-1} -0.06 {\tilde \xi_2}^{-2},&\;\; \tilde \xi_1 \ll1,\;\tilde\xi_2 \gg1,\\
                    0.06 \big({\tilde \xi_1}^{-2} - {\tilde \xi_2}^{-2}\big), &\;\; \tilde \xi_1,\tilde \xi_2 \gg1.
                   \end{cases}
\label{intdetadxi}
}
The constant term in the second line comes from integrating across the crossover regime around $\tilde \xi=1$ where the power-law dependence on $\tilde \xi$ changes. Choosing an arbitrary lower bound $\tilde \xi_0$, at which $\tilde\Lambda^{-1}\ll\tilde \xi_0\ll1$, we obtain
\al{
&\frac{\etaeff(L,\xi) - \etaeff(\xi_0)}{\eta_0} =  \frac{\ell}{(2 \pi)^2  L} \times \nl
&		   \begin{cases}
                    0.37\big(L/\xi_0 - L/\xi\big),&\mbox{for}\quad \xi\ll L,\\
                    -1.359 + 0.37 L/\xi_0 -0.06 L^2/\xi^2&\mbox{for}\quad \xi\gg L,
                   \end{cases}
}
from which we can also extract the bulk viscosity $\etaeff^{\textnormal{bulk}}$ at finite $\xi$,
\al{
\etaeff^{\textnormal{bulk}}(\xi) &\equiv \etaeff(L\to \infty,\xi) \nl
&= \etaeff(\xi_0) + {\eta_0 \frac{0.37\ell }{(2 \pi)^2 \xi}} \left(\frac \xi {\xi_0} -1\right).
}
In particular,  the {\it difference} between the bulk viscosity and that of a system where $\xi \gg L$ can now be expressed as 
\al{
\frac{\etaeff^{\textnormal{bulk}}(\xi)-\etaeff(\xi\gg L)}{\eta_0} = 
0.034 \,\tilde{\ell}\; Y(\tilde\xi)
\label{etaeffminbulk}
}
where $\tilde \ell = \ell/L$, and 
{
\al{
Y(\tilde\xi) = 1 - 0.27 \tilde \xi ^{-1} + 0.04 \tilde \xi ^{-2} +\mathcal{O}(\tilde \xi ^{-3}).
}
}
As long as $\xi \ll L$, contributions from the lower integration bound $\tilde \xi_0$ cancel in the subtraction. Equation \eqref{etaeffminbulk} is our main result: The difference considered in Eq.~\eref{etaeffminbulk}, normalized by $\eta_0$, is proportional to the ratio of the dynamical length scale {$\ell$} and the film thickness $L$, and diverges in the limit $L\to 0$ (bearing in mind that $L$ should anyhow be much larger than the microscopic length scale $\Lambda^{-1}$). These prefactors are multiplied by a scaling function {$Y(\tilde\xi)$, which is a function of $\tilde\xi$}, thus displaying universal properties.  
 
Equation \eqref{etaeffminbulk} can be tested experimentally or in non-equilibrium molecular dynamics simulations such as those in Ref.~\cite{roy2016}, if the viscosity can be measured at different values of $\xi/L$.

\section{Discussion and Outlook}
\label{discussionsect}

We have studied a confined, correlated fluid system driven out of equilibrium.  Through linear response theory for inhomogeneous shear, the self-consistent equations for fluid velocity were shown to take the form of a Stokes equation with a non-local viscosity. The effective viscosity of the system, defined in terms of the total stress in the steady state (or, equivalently, of the force necessary to shear the system), is found to be cutoff-dependent in general, as in the case of the corresponding bulk system. Relevant (finite) quantities were identified by taking the derivative of this effective viscosity with respect to correlation length $\xi$. This shows that the change of the viscosity with $\xi$ is independent of microscopic details, and as such is expected to be displayed universally. 

Specifically, we studied the case of a non-conserved order parameter coupled to a flowing medium, which could for instance describe systems with magnetic colloids where criticality is triggered magnetically. An interesting future perspective is the case of a conserved order parameter coupled to a flowing medium (Model H), which describes a binary fluid mixture. This analysis, left for future work, is more involved due to the added complexity in Model B dynamics in confined geometries \cite{diehljanssen1992,diehlwichmann1995,deangopinathan2009JStatMech}. Such an analysis would provide further insight into the relevant dynamical universality classes. Other avenues to explore include the incorporation of an external field and mixed or inhomogeneous boundary conditions for the film. Regarding response theory, an interesting question is whether an extension beyond the linear regime is possible. Such studies have been done for strong shear in bulk systems  \cite{fuchs} using a transient dynamics approach. Future work may also study the present model H(A) via renormalization group \cite{hohenberg,onukibook}, or consider a more microscopic theory \cite{kruegerdean2017}.

\begin{acknowledgments}
We thank G.~Bimonte, T.~Emig, N.~Graham,  R.~L.~Jaffe and M.~F.~Maghrebi for discussions {and especially M. Kardar for many useful suggestions and comments}. This work was supported by MIT-Germany Seed Fund Grant No.~2746830.  Ma.Kr.~and C.M.R.~are supported by Deutsche Forschungsgemeinschaft (DFG) Grant No.~KR 3844/2-1.
\end{acknowledgments}

\appendix

\section{Potential conditions in Model H}
\label{potcondsect}

In this Appendix we show that the potential conditions fix the form of the reversible force density induced by correlations of $\phi$, and demonstrate that these conditions hold irrespective of whether the dynamics of $\phi$ is purely dissipative or conserved. 

We condense the coupled Langevin equations \eref{langevinphi} and \eref{langevinv} as follows:
\al{
\partial_t \left( \begin{matrix}
 \phi \\
 \rho \bm v
 \end{matrix} \right)
 = \left( \begin{matrix}
 f_\phi \\
 (\fv)_\perp
 \end{matrix} \right)
 +
  \left( \begin{matrix}
 - \hat \mu \frac{\delta \mathcal H }{\delta \phi(\bm r, t)} \\
 \eta_0 \nabla^2 \frac{\delta \mathcal H}{\delta \bm v(\bm r, t)} 
 \end{matrix} \right)
 +
  \left( \begin{matrix}
 \theta \\ \bm \zeta
 \end{matrix} \right), \cm
 \label{ModHLangevin}
}
where $\mathcal H = \mathcal H[\phi] + \frac 1 2 \int d\bm r \rho \bm v^2 \cm$ is the full Hamiltonian, and we write the coupling terms as
\al{
f_\phi & =-  \bm \nabla \cdot (\phi \bm v)= - \bm v \cdot \bm \nabla \phi \qquad (\bm \nabla\cdot\bm v = 0),\nl
\bm f_{\bm v} &= - \bm \nabla \cdot T^\phi . \cm
}
The corresponding Fokker-Planck equation for the joint probability functional $\mathcal P[\phi, \bm v](t)$ is then  \cite{zwanzigbook,onukibook}
\al{
\partial_t \mathcal P = \hat L \mathcal P \equiv \big[ \hat L_e^\phi + \hat L_e^{\bm v} + \delta \hat L^\phi + \delta \hat L^{\bm v}  \big] \mathcal P.
\label{FPModH}
}
The ``diffusion'' terms (with irreversible contributions) are
\al{
\hat L_e^\phi\mathcal  P &=  \int d\bm r \frac{\delta}{\delta \phi} \hat \mu \big[ \frac 1 \beta \frac{\delta}{\delta \phi}+ \frac{\delta}{\delta \phi} \mathcal H \big] \mathcal P\cm\nl
\hat L_e^{\bm v} \mathcal P &= - \int d\bm r \frac{\delta}{\delta \bm v} \eta_0 \nabla ^2 \big[ \frac 1 \beta \frac{\delta}{\delta \bm v}+ \frac{\delta}{\delta \bm v}  \mathcal H \big]\mathcal  P,\cm
\label{diffterms}
}
and the ``streaming'' terms (with reversible contributions) are
\al{
\delta \hat L^\phi \mathcal P &= - \int d\bm r \frac{\delta}{\delta \phi} \big( \fp \mathcal P \big), 
\cm\nl 
\delta \hat L^{\bm v} \mathcal P &= - \int d\bm r \frac{\delta}{\delta \bm v} \big( \fv\mathcal  P \big). \cm
\label{straemterms}
}
The equilibrium probability distribution \eref{Peq} must lie in the null space of $\hat L$. (Note that $\hat L_e^{\bm v}\mathcal P_e = \hat L^{\phi}_e\mathcal P_e = 0$ is trivially satisfied.) This gives rise to the potential conditions on the streaming terms, i.e.,
\al{
\int d\bm r \Big[ \fp \frac{\delta}{\delta \phi} \mathcal H+ \fv \frac{\delta}{\delta \bm v} \mathcal H \Big] = \frac 1 \beta \int d\bm r \Big[\frac{\delta}{\delta \phi} \fp + \frac{\delta}{\delta \bm v} \fv \Big].\cm
\label{PotCondModH}
}
The second term on the r.h.s. of Eq.~\eref{PotCondModH} vanishes since the reversible force density induced by $\phi$ %$\phi \nabla \frac{\delta}{\delta \phi} \beta \mathcal H$ 
is independent of $\bm v$. The first term, instead, is $\int d\bm r  \frac{\delta}{\delta \phi} \fp = -\int d\bm r \frac{\delta}{\delta \phi} [  \bm v \cdot \bm \nabla \phi ].$ In general, $\frac{\delta}{\delta \phi(\bm r)} \bm \nabla_{\bm r'} \phi(\bm r') = - \bm \nabla \delta (\bm r - \bm r').$ Therefore, since $- \bm \nabla \delta (\bm 0) = 0$ (justified by considering the delta function as a limit of a Gaussian  \cite{deangopinathan2010PRE}), the first term is also zero. Accordingly, Eq.~\eref{PotCondModH} implies that $\int d\bm r [ -\bm  \nabla \cdot (\bm v  \phi)  \frac{\delta}{\delta \phi} \mathcal H + \fv \cdot \rho \bm v ]=0$. Since $\bm \nabla \cdot \bm v=0$, after integration by parts, this condition can be written as $\int d\bm r \; \bm v \cdot  [ -\phi \bm \nabla  \frac{\delta \mathcal H}{\delta \phi}  + \rho \fv ]=0$, which must hold for any $\bm v$. We conclude that the advection term $\fp$ therefore fixes the reversible force density $\fv$ to be
\al{
\fv& = -\phi \bm \nabla \frac{\delta}{\delta \phi} \mathcal H .\cm
\label{fixf}
}
As discussed in Sec.~\ref{fsect}, $\fv$ may also be viewed as the divergence of the stress tensor $T^\phi$ induced by $\phi$, i.e., 
$-\phi \bm \nabla \frac{\delta}{\delta \phi} \mathcal H = -\bm \nabla \cdot T^\phi$.

Importantly, the potential conditions are satisfied for \textit{both} Model H and our Model H(A), since compliance of the diffusion terms with the fluctuation-dissipation theorem ensures correct relaxation irrespective of conservation laws.

\section{Details of stress tensor calculation}
\label{FTappend}

\subsection{Green's functions}

For the Gaussian Hamiltonian \eref{Hgauss},
\al{
\frac{\delta \mathcal H}{\delta \phi} = -\nabla^2 \phi(\bm r,t)+\frac 1 {\xi^2}\phi(\bm r,t).
\label{dHdphi}
}
Correspondingly, for $\bm v = 0$ in the Langevin equation \eref{langevinphi} (the equilibrium case),  $\phi$ can be written in terms of Green's functions, 
\al{
\phi(\bm r,t)  = \int_V d\bm r' \int_{-\infty}^{\infty} dt' \;G(\bm r,t;\bm r',t') \theta(\bm r',t')
}
where $V$ is the volume of the system, and
\al{
\Big[\partial_t - \mu (\nabla^2_r - \frac 1 {\xi^2})\Big] G(\bm r,t;\bm r',t') = \delta^{(3)}(\bm r - \bm r') \delta(t-t').
\label{GFDE}
}
$G$ must then be determined subject to boundary conditions and geometry, allowing for computation of two-point correlation functions of $\phi$ as required for the linear response calculation. This is discussed in Appendix \ref{corrappend}. 

\subsection{Equilibrium field correlators}
\label{corrappend}

According to equations (10), (11) and (12) of Ref.~ \cite{hanke}, the general correlator can be written as
\al{
\ave{\phi&(\bm r,t)\phi(\bm r',t')} \nl
&  =  2\mu k_B T\int d\Gamma e^{-i\omega(t-t')} e^{i \bm p\cdot(\bm r_\parallel-\bm r_\parallel')} I(z,z'),
\label{generalcorr}
}
where $\int d\Gamma \equiv \int \frac{d^2 p}{(2\pi)^2} \int_{-\infty}^\infty \frac{d\omega}{2\pi}$. (Recall here the decomposition $\bm r = (z,\bm r_\parallel)$ and similarly for $\bm r'$.) Here we define
\al{
I(z,z') = \int_{\D} d\zeta\; g(z,\zeta;\omega,\bm p)  g^*(z',\zeta;\omega,\bm p)
\label{Idef}
}
in terms of the dynamic Green's $g$ function for Model A with Dirichlet boundary conditions in 
a film geometry $z \in \mathcal D= [-L/2,L/2]$. (Here $g$ is the Green's function in temporal Fourier space as well as spatial Fourier space for the parallel coordinates --- see Eqs.~\eref{generalcorr} and \eref{Idef}.) With the Hamiltonian \eref{Hgauss}, we see from the Green's function equation \eref{GFDE} that \cite{hanke}
\al{
g(z,z';\omega,\bm p)= 
    \begin{cases}
    \frac{
    \sinh [Q (z+\frac L 2)]\sinh [Q (\frac L 2 - z')]}{ \mu Q \sinh(Q L)},\; z<z', \\
    \; \\
    \frac{
    \sinh [Q (z+\frac L 2)]\sinh [Q (\frac L 2 - z')]}{  \mu Q \sinh(Q L)},\; z\geq z',
    \end{cases}
\label{gdeffilm}
}
with 
\al{
Q = \sqrt{p^2 + \frac 1{\xi^2}-i \frac \omega \mu }.
\label{Qdefappend}
} 
(In Ref.~ \cite{hanke} the case $\xi = \infty$ is considered; the modification to finite correlation length is trivially obtained by considering Eqs.~\eref{Hgauss}, \eref{dHdphi} and \eref{GFDE} by replacing $p^2$ with $p^2+1/\xi^2$.) 

Since $g$ is an even function of $p = |\bm p|$, complex conjugation in Eq.~\eref{Idef} implies replacing $Q$ with its complex conjugate, $P =Q^* = \sqrt{p^2 + \frac 1{\xi^2}+ i\frac \omega \mu}$. Explicitly,
\al{
I(z,z') = {\scriptstyle \begin{cases}
 \frac {P \text{csch}(L Q) \sinh (Q z) \sinh [Q(L-z')]}{\mu^2 PQ(P^2-Q^2)} - c.c. %Q \text{csch}(L P) \sinh (P z) \sinh [P(L - z')]-
 , & z<z' \\
 \; \\
 \frac{P \text{csch}(L Q) \sinh (Q z') \sinh [Q(L-z)]}{\mu^2{PQ(P^2-Q^2)}} -c.c., & z\geq z'.
    \end{cases}
    }
    \label{Ifilmdef}
}
The result is proportional to the imaginary part of $g$ --- this essentially encodes the fluctuation-dissipation theorem  \cite{gambassi2006}.

\subsection{Stress tensor autocorrelator}
\label{stresstenappend}
To compute $\czzp$ as defined in Eq.~\eref{czzpdef} we need to determine the correlator
\al{
\ave{\Delta \txz &(z,t)\;\Delta \txz(z',t')} \nl
&= \hat I\;\ave{\Delta \txz (\bm r,t)\;\Delta \txz(\bm r',t')} \nl
&= \hat I \hat L \hat D
\;\big\langle \phi(\bm r_1,t)\phi({\bm r_2},t)\phi(\bm r_1',t')\phi({\bm r_2}',t')\big\rangle,
\label{corr4}
}
where we have introduced the operators
\al{
\hat I &\equiv \int_0^\infty d(t-t')\int d \bm r_\parallel , \nl
\hat L &\equiv \lim_{\bm r_1, {\bm r_2} \rightarrow \bm r} \;\lim_{\bm r_1', {\bm r_2'} \rightarrow \bm r'} \quad \textnormal{and} \nl
\hat D &\equiv \partial_{x_1} \partial_{z_2} \partial_{x_1'} \partial_{z_2'}.% {\bm r_2}_\parallel \bm r_{2\parallel}
}
The four-point correlator in Eq.~\eref{corr4} can be Wick-contracted; we use the short-hand notation
\al{
\ave{1234}=\ave{12}\ave{34} + \ave{13}\ave{24} + \ave{14}\ave{23}.
}
The contribution
\al{
\hat L \hat D \ave{12}\ave{34} &= \ave{ \txz }_e
}
cancels with those that are subtracted in $\Delta \txz$. From the remaining two contributions we obtain
\al{
&\czzp  = \hat I \hat L \nl
& 
 \Bigg[  \Big\langle \partial_{x_1}\phi(\bm r_1,t) \partial_{x_1'}\phi(\bm r_1',t') \Big\rangle 
\Big\langle \partial_{ z_2}\phi({\bm r_2},t)\partial_{ z_2'}\phi({\bm r_2}',t') \Big\rangle +\nl
& \Big\langle \partial_{x_1}\phi(\bm r_1,t)\partial_{ z_2'}\phi({\bm r_2}',t') \Big\rangle
\Big\langle \partial_{z_2}\phi({\bm r_2},t)\partial_{x_1'}\phi(\bm r_1',t') \Big\rangle\Bigg].
\label{Tfluct}
}
For the first term of Eq.~\eref{Tfluct} we find from Eq.~\eref{generalcorr} that
\al{
&\big\langle \phi(\bm r_1,t) \phi(\bm r_1',t') \big\rangle \big\langle \phi({\bm r_2},t)\phi({\bm r_2}',t') \big\rangle =\nl
&\quad(2 \mu k_B T)^2 \int d\Gamma_1 \int d\Gamma_2 \;e^{-i\omega_1(t-t')} e^{i \bm p_1\cdot(\bm r_{1\parallel}-\bm r_{1\parallel}')}\nl
& \qquad e^{-i\omega_2(t-t')} e^{i {\bm p_2}\cdot({\bm r}_{2\parallel}-{\bm r}_{2\parallel}')}I(z_1,z_1') I(z_2, z_2 ').
}
The differential operator $\partial_{x_1} \partial_{x_1'}$ brings down $(i {p_1}_x)(-i{p_1}_x) = {p_1}_x^2$. The limits $\lim_{\bm r_1\rightarrow \bm r}$ and $\lim_{\bm r_1'\rightarrow \bm r'}$ change the term 
$e^{i \bm p_1\cdot(\bm r_{1\parallel}-\bm r_{1\parallel}')}$ and set $z_1\rightarrow z$, $z_1'\rightarrow z'$. The limits $\lim_{{\bm r_2} \rightarrow \bm r}$ and $\lim_{{\bm r_2'} \rightarrow \bm r'}$ change 
$e^{i {\bm p_2}\cdot({\bm r}_{2\parallel}-{\bm r}_{2\parallel}')}$ and change where the $\partial_{ z_2}$ and  $\partial_{z_2'}$ derivatives are evaluated. $\int d\bm r_\parallel$ gives $\textcolor{black}{(2\pi)^2} \delta(\bm p_1 + {\bm p_2})$ and $\int \frac{d^2 p_2}{\textcolor{black}{(2\pi)^2}}$ then sets ${\bm p_2}=-\bm p_1 $ throughout.
Lastly, it suffices to consider the real part of the time integral,
$ \Re \big[ \int_0^\infty d(t-t')\; e^{-i(t-t')(\omega_1 + \omega_2)} \big] = \int_0^\infty d\tau\; \cos[\tau(\omega_1 + \omega_2)] = \textcolor{black}{\frac 1 2} \textcolor{black}{(2\pi)} \delta(\omega_1 + \omega_2).$
What remains for this term is 
\al{
&\hat I \hat L \hat D \Big\langle \partial_{x_1}\phi(\bm r_1,t) \partial_{x_1'}\phi(\bm r_1',t') \Big\rangle 
\Big\langle \partial_{ z_2}\phi({\bm r_2},t)\partial_{ z_2'}\phi({\bm r_2}',t') \Big\rangle
\nl
&={\frac 1 2} (2 \mu k_B T)^2 
\int d\Gamma \;p_x^2 \;I(z,z') \;\partial_{z} \partial_{z'} I^*( z,  z').
\label{int1}
}

For the {second term} of Eq.~\eref{Tfluct} the analysis is similar, except that a relative minus sign arises since the derivatives act on different terms.
The result is
\al{
&\hat I \hat L \hat D \Big(\langle \partial_{x_1}\phi(\bm r_1,t)\partial_{ z_2'}\phi({\bm r_2}',t') \Big\rangle
\Big\langle \partial_{z_2}\phi({\bm r_2},t)\partial_{x_1'}\phi(\bm r_1',t') \Big\rangle \nl
&=-\textcolor{black}{\frac 1 2} (2 \mu k_B T)^2 
\int d\Gamma \;p_x^2 \;\big[\partial_{z} I(z,z ')\big]\big[\partial_{z'} I^*(z, z')\big].
\label{int2}
}

Note the relative minus sign between the integrands in the first and second term of Eq.~\eref{Tfluct}. Combining these expressions gives
\al{
\czzp = \frac{( k_B T)^2}{(2\pi )^2}\int_{0}^\infty dp \;\; f(z,z',p,L,\xi),
\label{czzpfilm}
}
where the integrand $f$ is
\begin{widetext}
\al{
f(z,z',p,L,\xi) &= \int d\omega\;\; \frac{p^3 \text{csch}(L P) \text{csch}(L Q) }{P Q \left(P^2-Q^2\right)^2}\nl
&\quad \times \Bigg[P \cosh \big( P (z+\frac L 2) \big) \sinh \big( Q (z+\frac L 2)\big)-Q \sinh \big( P (z+\frac L 2)\big) \cosh \big( Q (z+\frac L 2)\big)\Bigg] \\
&\quad \times \Bigg[P \cosh \big( P (\frac L 2-z')\big) \sinh\big( Q (\frac L 2-z')\big)-Q \sinh \big( P (\frac L 2-z')\big) \cosh \big( Q ( \frac L 2-z')\big)\Bigg], \nonumber
}
\end{widetext}
with 
\al{
Q(p,\omega,\xi) &= \sqrt{p^2 + 1/\xi^2 - i \omega/\mu} \quad\textnormal{and} \nl
P(p,\omega,\xi) &=\sqrt{p^2 + 1/\xi^2 + i \omega/\mu}= Q^*.
\label{Qdef}
}
To obtain a finite result, the $p$ integral in Eq.~\eref{czzpfilm} requires an upper cutoff $\Lambda = 1/a$, where $a$ is a molecular microscopic length scale.

\begin{widetext}
\onecolumngrid

\begin{figure}[t]
 \centering
 \includegraphics[width=0.65\textwidth]{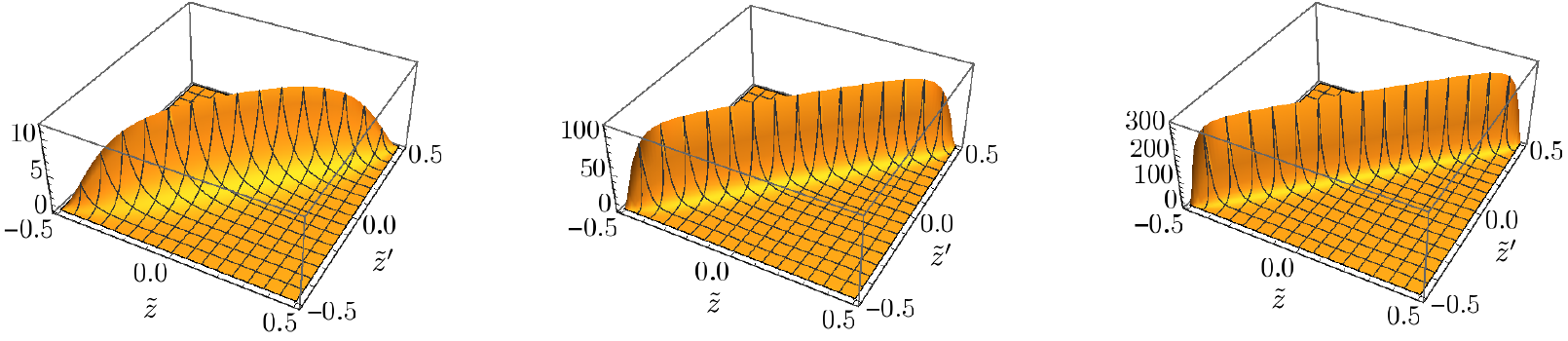}
 \caption{The rescaled part $\czzplr$ of the stress-stress autocorrelation function $\czzpl$ in Eq.~\eqref{czzpdef}, which is also related to the non-local viscosity, $\eta(z,z') = \frac {1} {k_B T} \czzpl$, as a function of rescaled coordinates $\tilde z$ and $\tilde z'$, ranging from $-1/2$ to $1/2$. Curves from the left to the right correspond to the rescaled cutoff $\tilde \Lambda=10,30,50$. The correlation length is set to $\tilde \xi = \infty$. }
 \label{fig:czzpLam}
\end{figure}

\end{widetext}

\twocolumngrid

\subsection{Rescaling for film}
\label{rescaleappend}

Consider Eq.~\eref{czzpfilm} with a cutoff $\Lambda$. We define rescaled coordinates $\tilde z = z/L$, $\tilde z' = z'/L$, $\tilde \xi = \xi/L$, $\tilde p = p L$, $\tilde \omega = \omega L^2/\mu$. In terms of these, we find from Eq.~\eref{Qdef} that
\al{
\tilde Q =Q(\tilde p,\tilde \omega,\tilde \xi) &= \sqrt{\tilde p^2 + 1/\tilde \xi^2 - i \tilde \omega} = L Q(p,\omega,\xi) \quad\textnormal{and} \nl
\tilde P = P(\tilde p,\tilde \omega,\tilde \xi) &=\sqrt{\tilde p^2 + 1/\tilde \xi^2 + i\tilde \omega} = L P(p,\omega,\xi).
\label{Qrdef}
}
Therefore
\al{
f(z,z',p,L,\xi) = L \tilde f(\tilde z,\tilde z'),
\label{ftilde}
}
where $\tilde f$ is obtained from $f$ by replacing $Q \to \tilde Q$, $P \to \tilde P$, $p \to \tilde p$, $z+ L/2 \to \tilde z + 1/2$ and $ L/2 - z' \to 1/2 - \tilde z'$.
What remains to be evaluated are the $p$ and $\omega$ integrals in Eq.~ \eref{czzpfilm}, wherein $dp = \frac 1 L d\tilde p$ and $d\omega = \frac \mu {L^2} d\tilde \omega$. Inserting this and Eq.~\eref{ftilde} into Eq.~\eref{czzpfilm} gives  Eq.~\eref{czzpdef}.

\subsection{Cutoff dependence of the stress-stress autocorrelation function $\czzp$}
\label{exampleappend}

\begin{figure}[t]
 \centering
 \includegraphics[width=0.45\textwidth]{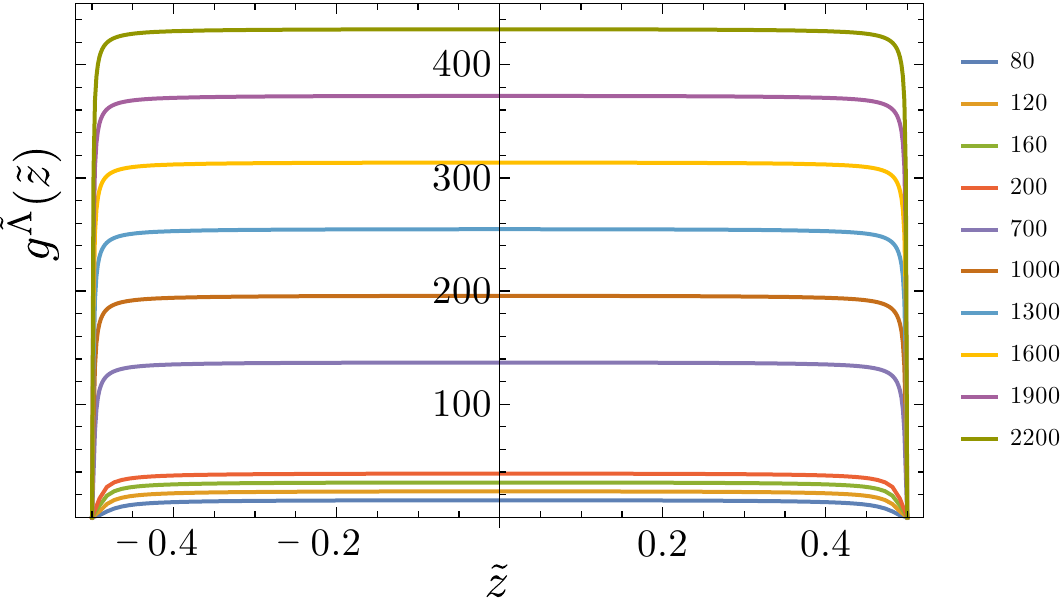}%{./Figs/g(z)_Lambda.pdf}
  \caption{The envelope function $g^{\tilde \Lambda}(\tilde z)$ --- see Eq.~\eqref{ansatz} --- for the various values of $\tilde \Lambda$ indicated in the legend ($\tilde \xi = \infty$). For $\tilde \Lambda\to\infty$, the form of Eq.~\eqref{gcases} is approached.} 
 \label{fig:g(z)}
\end{figure}

\begin{figure}[t]
 \centering
 \includegraphics[width=0.43\textwidth]{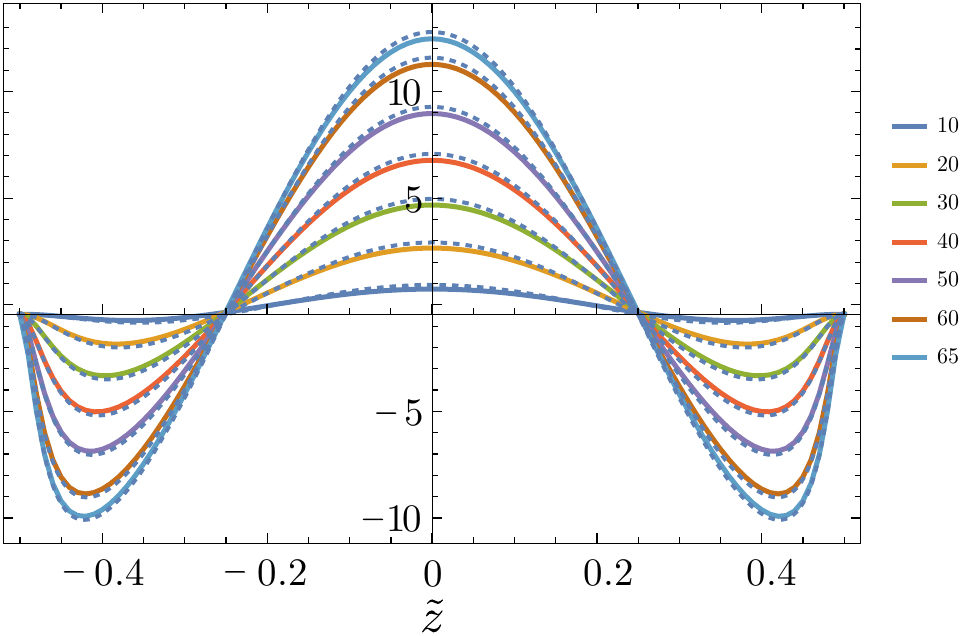}
  \caption{$\int_{-1/2}^{1/2} d\tilde z' \;\cos(2 \pi \tilde z')\czzplxiinfr$ (solid lines) and the envelope Ansatz in Eq.~\eref{ansatz} (dotted) for various values of $\tilde \Lambda$, as indicated in the legend, with $\tilde \xi = \infty$. }
 \label{fig:cos}
\end{figure}

In Sec.~\ref{corrsect} we remarked that the function $\czzp$, which is the kernel of the linear response calculation and may be viewed as the non-local viscosity in the Stokes equation, becomes increasingly local as $\Lambda$ is increased. This behaviour is shown in Fig.~\ref{fig:czzpLam}, and motivates the envelope Ansatz in Eqs.~\eref{ansatz} and \eref{gcases}, as illustrated in Fig.~\ref{fig:g(z)}. As an example of the accuracy of this Ansatz, Fig.~\ref{fig:cos} compares the actual integral
$\int_{-1/2}^{1/2} d\tilde z' \;\cos(2 \pi \tilde z')\czzplxiinfr$
with the Ansatz
$g^{\tilde \Lambda}(\tilde z) \cos(2 \pi \tilde z')$, showing that their agreement improves as $\tilde \Lambda \to \infty$.

\section{Cutoff dependence of the shear rate, total stress and effective viscosity}
\label{critappend}
\begin{figure}[t]
 \begin{center}
 \includegraphics[width=0.45\textwidth]{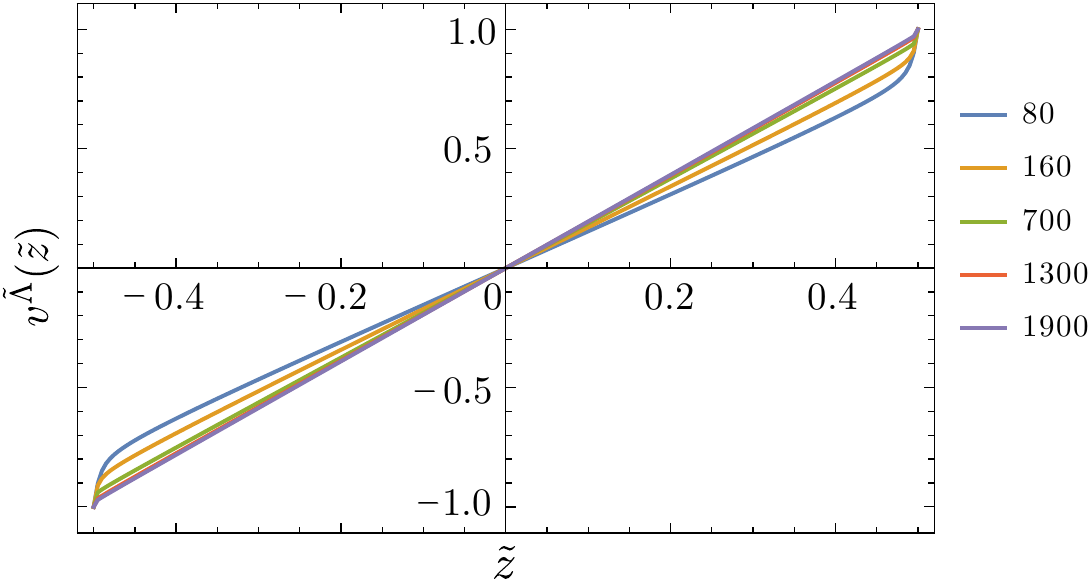}
 \caption{Velocity profile $v^{\tilde \Lambda}(\tilde z)$ for the confined system, in units of the boundary velocity $v^*$, computed with $\alpha=1$ for various values of the cutoff, $\tilde \Lambda=80,160,700,1300,1900$ ($\tilde\xi = \infty$). As $\tilde \Lambda\to\infty$, the dependence on $\alpha$ disappears. (The latter also follows by considering Eq.~\eref{eq:sr-largeL} for large cutoffs.)}
 \label{fig:v(z)}
 \end{center}
\end{figure}

For asymptotically large values of $\Lambda$, the non-local viscosity kernel $\eta(z,z')$ (recall Eq.~\eref{nonlinstokes1}) becomes increasingly local, and the effective viscosity diverges and approaches the bulk result of the same model  \cite{onukibook}. For a finite value of $\Lambda$, the velocity profile $v^{\tilde \Lambda}$, as seen in Fig.~\ref{fig:v(z)} for $\xi = \infty$, shows a deviation from the simple shear profile, and the effective viscosity scales as $\etaeff\sim 1/L$. In order to get rid of this dependence on $\Lambda$, and to highlight those features of the confined system which depend primarily on the confinement but not on the microscopic details, 
we consider in Sec.~ \ref{corrsect} the dependence of these quantities on  the correlation length $\xi$, which is demonstrated to be independent of $\Lambda$, and is thus expected to describe  universal behavior. 

It is clear from Fig.~\ref{fig:v(z)} that the shear rate approaches simple shear, $s(z)\to s_0 = 2v^*/L$, as $\tilde \Lambda \to \infty$. Correspondingly the velocity profile becomes asymptotically linear. (Note that the results for both these figures were obtained through the Ansatz in Eq.~\eref{ansatz}, but agree well with actual iterative calculations using the full $\tilde z$ integral with  $\czzplr$  for large $\tilde \Lambda$.) This also occurs when $\xi \neq \infty$, as long as $a = \Lambda^{-1} \ll \xi$.

\begin{figure}[t]
 \begin{center}
 \includegraphics[width=0.45\textwidth]{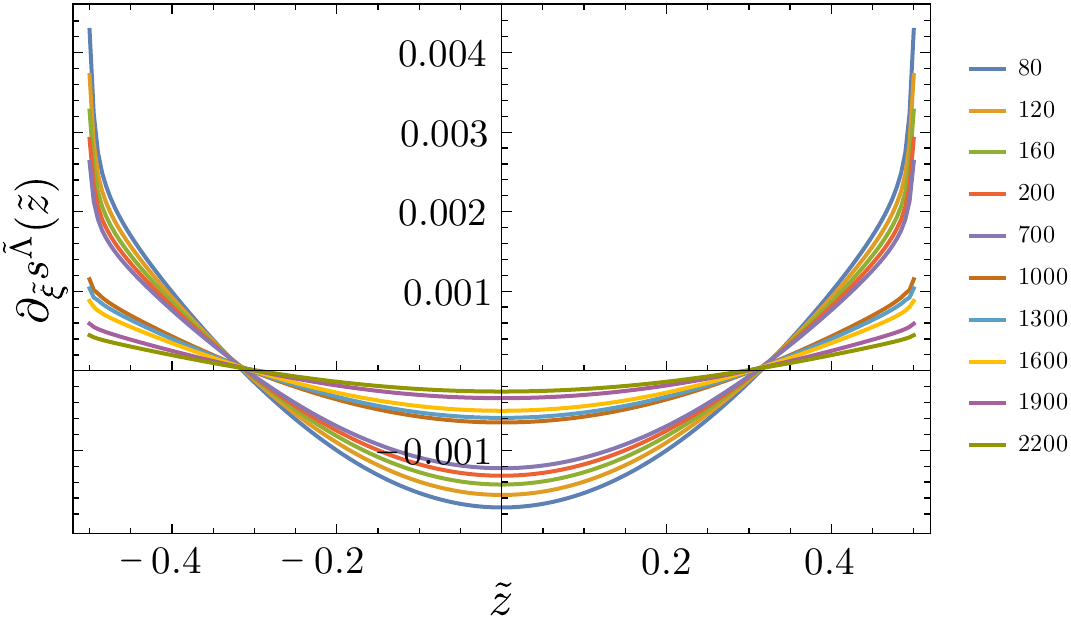}
  \caption{$\partial_\xi s(\tilde z)$ in units $s_0\alpha {\tilde \xi}^{-3}$ with $\alpha = 1$, shown on the range $\tilde z \in [-0.48,0.48]$. The integral of this function approaches zero as $\tilde\Lambda\to\infty$.}
  \label{fig:dxisZoom}
 \end{center}
\end{figure}

In Sec.~ \ref{corrsect} we computed the self-consistent solution for $\partial_{\tilde \xi} s(\tilde z)$; this is shown for various (large) $\tilde \Lambda$ in Fig.~\ref{fig:dxisZoom} for $\xi = \infty$. Clearly $\partial_{\tilde \xi} s(\tilde z)$ diverges at the boundaries; this divergence comes closer to the boundaries as the cutoff is increased. In the limit $\tilde \Lambda\to\infty$, we have also checked that $\partial_{\tilde \xi} s(0) \sim1/\tilde \Lambda \to 0 $, i.e., the derivative w.r.t. $\tilde \xi$ of the shear rate (and therefore also of the velocity profile) becomes independent of $\tilde \xi$ near the critical point for large $\tilde \Lambda$.

\clearpage

\end{document}